%% file: warpU_JASA_revision.tex
\documentclass[12pt]{article}
\usepackage{amsmath}
\usepackage{graphicx}
\usepackage{enumerate}
\usepackage{natbib}
\usepackage{url} 

\newcommand{\blind}{1}

\usepackage[hidelinks,citecolor=blue,colorlinks=true,linkcolor=blue]{hyperref}
\onecolumn 

\graphicspath{{Fig/}}

\usepackage{booktabs}

\usepackage{algorithm}
\usepackage{amsmath}
\usepackage{graphicx}
\usepackage{url} 
\allowdisplaybreaks
\usepackage[utf8]{inputenc} 
\usepackage{amsfonts}       
\usepackage{nicefrac}       
\usepackage{microtype}      
\usepackage{lipsum}
\usepackage{comment}
\usepackage{multirow}
\usepackage{algorithmic}
\usepackage[textwidth=\marginparwidth]{todonotes}
\usepackage{xcolor}

\usepackage{subcaption}
\usepackage{afterpage}

\usepackage{amsfonts}       
\usepackage{amsthm}
\theoremstyle{thmstyleone}%

\newtheorem{theorem}{Theorem}

\usepackage{graphicx}
\usepackage{amsmath} 

\input{notation.tex}

\usepackage{soul}
\usepackage[normalem]{ulem}

\usepackage{xr}

\makeatletter
\newcommand*{\addFileDependency}[1]{
	\typeout{(#1)}
	\@addtofilelist{#1}
	\IfFileExists{#1}{}{\typeout{No file #1.}}
}
\makeatother

\usepackage{authblk}

\begin{document}

\def\spacingset#1{\renewcommand{\baselinestretch}%
{#1}\small\normalsize} \spacingset{1}


\date{}
\if1\blind
{
  \title{\bf Channelling  Multimodality Through a Unimodalizing Transport: Warp-U Sampler and Stochastic Bridge Sampling Estimator}
  
\author[1]{\normalsize Fei Ding\footnote{Equal Contribution.}}
\newcommand\CoAuthorMark{\footnotemark[\arabic{footnote}]}
\author[2]{Shiyuan He\protect\CoAuthorMark}
\author[1]{David E. Jones\protect\CoAuthorMark}
\author[3]{Xiao-Li Meng}   
\affil[1]{\normalsize Department of Statistics, Texas A\&M University, College Station, TX, USA.} 
\affil[2]{School of Mathematics and Statistics, Beijing Technology and Business University, Beijing, China.} 
\affil[3]{Department of Statistics, Harvard University, Cambridge, MA, USA.}  

  \maketitle
} \fi

\if0\blind
{
  \bigskip
  \bigskip
  \bigskip
  \begin{center}
  \spacingset{1.5} 
    {\LARGE\bf Channelling  Multimodality Through a Unimodalizing Transport: Warp-U Sampler and Stochastic Bridge Sampling Estimator}
\end{center}
} \fi
\spacingset{1.3} 
\medskip
\begin{abstract}
Monte Carlo integration is a powerful tool for scientific and statistical computation, but faces significant challenges when the integrand is a multi-modal distribution, even when the mode locations are known. This work introduces novel Monte Carlo sampling and integration estimation strategies for the multi-modal context by leveraging a generalized version of the stochastic Warp-U transformation \citep{wang2022warp}. We propose two flexible classes of Warp-U transformations, one based on a general location-scale-skew mixture model and a second using neural ordinary differential equations. We develop an efficient sampling strategy called \textit{Warp-U sampling}, which applies a Warp-U transformation to map a multi-modal density into a uni-modal one, then inverts the transformation with injected stochasticity. In high dimensions, our approach relies on information about the mode locations, but requires minimal tuning and demonstrates better mixing properties than conventional methods with identical mode information. To improve normalizing constant estimation once samples are obtained, we propose a {\it stochastic Warp-U bridge sampling estimator}, which we demonstrate has higher asymptotic precision per CPU second compared to the original approach proposed by \citet{wang2022warp}. We also establish the ergodicity of our sampling algorithm. The effectiveness and current limitations of our methods are illustrated through simulation studies and an application to exoplanet detection.
\end{abstract}

\noindent%

\spacingset{1.7}

\section{Monte Carlo Integrations}

\subsection{Computing normalizing constants --- How hard can it be?}

Computing or estimating normalizing constants may sound like a routine homework exercise, 
but it is a surprisingly wide-ranging and hard problem in statistics and more broadly in many scientific applications, from computing free energy in physics \citep[see][]{bennett1976efficient} to cognitive studies in psychology \citep[see][]{gronau2019simple,gronau2020computing}. Bayesian evidence ratios (a.k.a., Bayes factors) are often computed via normalizing constant estimation techniques, and are widely used for hypothesis testing and model selection in many scientific fields.
For example,  \cite{nelson2018quantifying} and \cite{pullen2014bayesian} discussed the computation of Bayesian evidence in the context of exoplanet detection and systems biology, respectively. 

Given the many scientific uses of normalizing constants, computationally and statistically efficient methods for estimating them are of high practical value, and  many powerful algorithms have been (re)invented, as we shall review briefly in Section~\ref{sec:litreview}.  However, some ubiquitous scenarios remain challenging, such as estimating normalizing constants for multi-modal target densities. For example, \cite{nelson2018quantifying} applied numerous strategies for estimating the Bayesian evidence for the presence of an exoplanet orbiting a star and obtained somewhat divergent estimates, even after substantial calibration efforts. 

For any absolutely integrable function $f$ we can write
\begin{equation}\label{eq:general}
	\int_\Theta f(\vtheta)\mud\vtheta = \int_\Theta f_+(\vtheta)\mud\vtheta - \int_\Theta f_-(\vtheta)\mud\vtheta,
\end{equation}
where $\mu$ is a baseline measure\footnote{See \cite{kong2003theory} for the essential role the baseline measure $\mu$ plays in forming an appropriate likelihood theory for Monte Carlo integration.}, and both $f_+=\max\{f(\vtheta), 0\}$ and $f_-=-\min\{f(\vtheta), 0\}$ are non-negative functions and hence can be viewed as unnormalized densities on $\Theta$. Consequently, although we will frame our approaches in the context of normalizing constants, they are more generally applicable because of identity \eqref{eq:general}, and because of other ways of making the integrand positive, e.g., using a lower bound on $f$ as a control variate.

We consider the following problem: for an unnormalized probability density $q$ with support $\Theta$, we seek to estimate the normalizing constant  $c = \int_\Theta q(\vtheta)\mud\vtheta$. Any Monte Carlo approach for this problem involves two key tasks: (i) obtaining samples $\vtheta_1,\vtheta_2, \cdots,\vtheta_n$ from $\target=q/c$ or some related density (or densities), and (ii) constructing an estimate of $c$ based on the Monte Carlo samples. These tasks may be performed sequentially or in combination, depending on the specific strategy. It is worth emphasizing that when $\target$ is multi-modal,  substantial inefficiencies can result from not addressing the difficulties created by multi-modality in both (i) and (ii) (on the other hand addressing multi-modality does not automatically guarantee  an efficient estimator of $c$). 

Computing normalizing constants is  exceedingly challenging and often requires methods tailored to the particular context. Indeed, a completely general and efficient method for performing integration seems impossible with contemporary software, hardware, and \textit{mindware}\footnote{A term coined by psychologist David Perkins, referring to the knowledge and problem solving techniques available to human minds.}. Our approaches add to the toolkit for handling multi-modality in Monte Carlo, but it would be unwise to promote them without understanding their pros and cons, just as with any methods proposed before or after. Indeed, we believe the great potential of the unimodalizng strategy is yet to be explored,  especially when blended with other powerful approaches such as unbiased Monte Carlo sampling \citep{jacob2020unbiased}.

\subsection{An Integrated Approach for Combating Multi-modality}
\label{sec:intro:contributions}

We propose a Warp-U Markov chain Monte Carlo (MCMC) sampler for effective sampling from mutli-modal densities, and a complementary strategy for estimating the corresponding target normalizing constant. The estimator is complementary in the sense that much of the necessary computation is completed during the sampling stage.  Our framework builds upon the work of \cite{wang2022warp}, which introduced the stochastic Warp-U transformation to convert multi-modal densities into approximately uni-modal ones and leveraged it in normalizing constant estimation. Specifically, we introduce (i) a more general class of Warp-U transformations, (ii) an MCMC sampler based on Warp-U transformations (whereas \cite{wang2022warp} focused on estimation {\it given} samples), and (iii) a more computationally efficient Warp-U bridge sampling estimator for normalizing constants. 

A key ingredient for an effective Warp-U transformation \citep{wang2022warp} is a reasonable mixture approximation of the target density. Warp-U bridge sampling estimators are asymptotically unbiased for any mixture approximation of the target, but their efficiencies depend on the degree of approximations.\footnote{Note that a well-constructed Warp-U bridge sampling estimator is more efficient than a Metropolis-Hastings algorithm using the same mixture approximation as a proposal density, see \citet{wang2022warp}.}
In Section~\ref{sec:extendedWarpU}, 
we introduce a class of Warp-U transformations  based on a general location-scale-skew family of mixture models.
Our class of transformations can effectively capture a wide range of local features of the target distribution (e.g., skewness and heavier tails), thereby enhancing Warp-U estimation efficiency.

In Section~\ref{sec:stochasticBridgeEstimation}, 
to improve upon the Warp-U bridge estimator (WB) for normalizing constants introduced  by  \cite{wang2022warp},
 we propose a \textit{stochastic} Warp-U bridge estimation  strategy (S-WB).  S-WB applies a divide-and-conquer approach to increase computational efficiency. For a given set of input samples, S-WB requires  fewer target distribution queries than WB, yet achieves comparable or even superior estimation accuracy.
 
Sections~\ref{sec:distpre} and \ref{section:sampling} develop an MCMC sampling algorithm, the {\it Warp-U sampler}, which alternately applies the stochastic forward Warp-U transformation $\sF_{\zeta}$ and its inverse $\sF_{\zeta}^{-1}$.  Since all local components of the target distribution are mapped to similar intermediate uni-modal densities, the subsequent mapping back to the target density tends to have a high probability of relocating a given draw to a different local component than the one originally sampled. This property makes the proposed procedure effective for multi-mode exploration, as initially demonstrated in the first version\footnote{ See Section~6 of the initial version at {\tt https://arxiv.org/pdf/1609.07690v1.pdf}, which was removed in the published version, following an editorial request to streamline \cite{wang2022warp}.} of \cite{wang2022warp}.  Again a reasonable mixture approximation to the target is crucial, which was one of the reasons we developed more general classes of Warp-U transformations. 

In Section~\ref{sec:neural}, we gain further flexibility by constructing non-linear transformations using neural ordinary difference equations (ODEs) \citep{chen2018neural}. These  non-linear Warp-U transformations have excellent sampling (and estimation) performance in our numerical studies, albeit at some initial cost of constructing the transformations.   Section~\ref{sec:theory} summarizes our theoretical contributions for analyzing the proposed estimators and samplers.  

We emphasize that, in very high-dimensional settings, currently we can replicate our sampler’s superior mixing properties only after the modes are approximately located, e.g., by an optimization algorithm with random initial values. Specifically, Section~\ref{sec:simulationstudies} empirically examines our integrated approach through simulation. Section~\ref{realdataanalysis} applies our methods to the estimation of Bayesian evidence for exoplanet discovery. The experiments demonstrate the proposed methods' potential for outperforming existing sampling and estimation strategies, which are applied together in various combinations, including combining parallel tempering with the Warp-U bridge estimator.  In order to report our main proposals as soon as possible, we defer the literature review to Section~\ref{sec:literatureReview}, where we also discuss the limitations of our approach and possible improvements and extensions.  Additional numerical results and technical proofs are given in the  Supplementary Material. 

\vspace{-5pt}
\section{Warp-U Sampling and Estimation Methods}\label{sec:sampling}

\vspace{-10pt}
\subsection{Bridge Sampling Estimation}\label{sec:review}

To provide the necessary notation and background, this sub-section briefly reviews the parts of \cite{meng1996simulating}, \cite{meng2002warp}, and \cite{wang2022warp} on which we build, before we present our main proposals in the rest of  Section~\ref{sec:sampling}. Let $q_1$ and $q_2$ denote unnormalized densities with unknown normalizing constants $c_1$ and $c_2$, respectively, and for simplicity we assume they share the  support $\Theta= \mathbb{R}^d$. We are interested in estimating   $r=c_1/c_2$ (e.g., a Bayes factor). For this scenario, bridge sampling \citep{bennett1976efficient,meng1996simulating} relies on the  identity: 
$$	r = \frac{c_1}{c_2} = \frac{\Expect_{p_2}[q_1(\vtheta)\alpha(\vtheta)]}{\Expect_{p_1}[q_2(\vtheta)\alpha(\vtheta)]},$$   where $\alpha$ is the bridge function (discussed below), and  $\Expect_{p_i}$ denotes an expectation with respect to $\vtheta\sim p_i=q_i/c_i$, for $i=1,2$. This identity leads to  the bridge sampling estimator:
\begin{equation}\label{calssical_Bridge_estimation}
	\hat{r}=\frac{n_2^{-1}\sum_{j=1}^{n_2}q_1(\vtheta_{2,j})\alpha(\vtheta_{2,j})}{n_1^{-1}\sum_{j=1}^{n_1}q_2(\vtheta_{1,j})\alpha(\vtheta_{1,j})},  
\end{equation}
where $\left\{\vtheta_{i,1},\dots,\vtheta_{i,n_i}\right\}$ are (possibly dependent) samples from $p_i$, and  $n_i$ is the number of samples from $q_i$, for $i=1,2$.  When the samples are independent, \cite{meng1996simulating} showed that the optimal  $\alpha$ is 
	$\alpha_r(\vtheta) \propto [n_1q_1(\vtheta) + rn_2q_2(\vtheta)]^{-1}$,
which yields the smallest asymptotic variance for the estimator $\hat{r}$.  The issue of the unknown $r$  in the optimal choice of $\alpha$  is addressed by  an iterative scheme $\{r^{(t)}, t=1, \ldots, \}$, where $r^{(t)}$ is given by \eqref{calssical_Bridge_estimation} with $\alpha = \alpha_{r^{(t-1)}}$.  \cite{meng1996simulating}   showed that $\lim_{t\rightarrow \infty} r^{(t)}$ maintains the asymptotic variance of the optimal bridge sampling estimator for $r$ that uses  the true value of $r$ in $\alpha_r$.\footnote{This turns out to be a consequence of the fact that the fix-point equation implied by \eqref{calssical_Bridge_estimation} and $\alpha_r(\vtheta)$
	 is equivalent to the score equation for the maximum likelihood estimator of $r$ from the likelihood theory (for Monte Carlo integration) as formulated in \cite{kong2003theory}, and hence the adaption does not lead to loss of information, at least asymptotically. }

When we only need to deal with a single  unnormalized density $q=c\target$,  bridge sampling is still applicable, and indeed often preferred. We can set $q_1=q$ in \eqref{calssical_Bridge_estimation}, and then  choose a ``pairing'' (and normalized) density $q_2=p_2$, such as a Normal distribution. 
The choice of $p_2$ is important because the asymptotic variance of $\hat{c}$ decreases as the separation between $p_1=\target$ and $p_2$ decreases. Their separation is measured by  the  harmonic divergence
$	H_A(p_1,p_2) = 1- \int  \big[\eta_1p_1^{-1}(\vtheta) + \eta_2p_2^{-1}(\vtheta)\big]^{-1} \mud\vtheta$,
where $\eta_i \propto n_i^{-1}$. Obviously, $p_2(\vtheta)$ should also be chosen to minimize computation in terms of both function evaluation and sampling. 

Given the above considerations, if $\target=q/c$ is multi-modal, it is natural to  choose $p_2$  to be a Gaussian mixture distribution $ \phi_{\text{mix}}$ approximating $\target$.    Standard bridge sampling estimation would proceed by applying~\eqref{calssical_Bridge_estimation} with densities  $q_1 = q$ (unnormalized) and $q_2 = \phi_{\text{mix}}$. The top left panel of Figure~\ref{fig:warpuexample} provides an illustrative example in which $\phi_{\text{mix}}$ (dashed curve) is a three component Gaussian mixture and roughly approximates $\pi$ (solid curve). In general, the quality of  the approximation affects the accuracy of the estimator $\hat{r}$ in~\eqref{calssical_Bridge_estimation}.

\citet{wang2022warp} proposed an improved bridge sampling estimator based on the idea of warp bridge sampling estimation \citep{meng2002warp}. Utilizing properties of $f$-divergences \citep{ali1966general}, where the Harmonic divergence is a special case, any transformation $\mathcal{F}$ satisfies the inequality $H_A(\pi, \phi_{\text{mix}}) \geq H_A(\mathcal{F}(\pi), \mathcal{F}(\phi_{\text{mix}}))$. Since the asymptotic variance of $\hat{r}$ decreases with the Harmonic divergence, transforming the densities while retaining the normalizing constant generally reduces variance. A well-chosen transformation $\mathcal{F}$ can significantly improve efficiency. \citet{wang2022warp} proposed a transformation known as the Warp-U transformation, using a Gaussian mixture distribution.

\vspace{-2pt}
\subsection{Location-Scale-Skew Warp-U Transformations}
\label{sec:extendedWarpU}

Building upon the work of \citet{wang2022warp}, we introduce a broader class of  transformations to enhance computational efficiency. Denote $\phi(\cdot; \vmu, \mSigma)$ as the  density  of  the  Gaussian distribution $\mathcal{N}_d(\vmu, \mSigma)$ with mean $\vmu$ and covariance matrix $\mSigma$,
and $\phi(\cdot)$ as the standard Gaussian density.
 Consider the following mixture distribution
\begin{equation} \label{eqn:infiniteMixMain}
	\phi_{\text{mix}}(\vtheta)
	=\sum_{k=1}^{K} 
	 w_k  \int \phi(\vtheta;\ \vmu_k + u \valpha_k, v\mSigma_k ) p(u,v | \eta_k) \mathrm{d} u\mathrm{d} v \equiv  \sum_{k=1}^{K} \phi^{(k)}(\vtheta),
\end{equation}
where $w_k$, $\vmu_k$, $\valpha_k$ and $\mSigma_k$ denote the mixture weight, mean vector, skewness vector and covariance matrix for the $k$th mixture component, respectively.  Additionally, $p(u,v | \eta_k)$
is a density  for $(u,v)$ with $u\ge 0$ and is  parameterized by $\eta_k$.   In~\eqref{eqn:infiniteMixMain} and throughout, we let $\phi^{(k)}$ be the $k$-th component of $\phi_{\text{mix}}$ including its  weight $w_{k}$, and  we will also use
$\mS_k := \mSigma_k^{1/2}$. 

The mixture distribution~\eqref{eqn:infiniteMixMain}  is general and includes many interesting  cases.
 When $\valpha_k=\vzero$
	and $ p(u,v | \eta_k)$ is a point mass at $(u,v)=(1,1)$ for all $k\in\{1,\ldots, K\}$,
the mixture distribution~\eqref{eqn:infiniteMixMain} becomes a Gaussian mixture distribution:
	\begin{equation}
	\phi_{\text{mix}}^{\rm gauss}(\vtheta) = \sum_{k=1}^{K} \phi^{(k)}(\vtheta)  = \sum_{k=1}^{K}
	w_k |\mS_k^{-1}|\phi(\mS_k^{-1}(\vtheta-\vmu_k)). \label{eqn:phimix:gauss}   
	\end{equation}
When $ p(u,v | \eta_k)$ is a degenerate  point mass for $v$  at $v=1$,  
a mixture component in~\eqref{eqn:infiniteMixMain} becomes a  skewed-Gaussian distribution \citep{azzalini2005skew, lin2019stein}.
For example, we can let $ p(u,v | \eta_k) \equiv p(u | \eta_k)$ to be a  Gamma distribution or a Gaussian distribution truncated to $u\in [0,\infty)$.  For the latter, it holds for a mixture component that
	\begin{equation}  \label{eqn:phimix:skew}   
	\footnotesize		\phi^{(k)}(\vtheta)  = w_k \int \phi(\vtheta;\ \vmu_k + |u| \valpha_k, \mSigma_k ) \phi(u) \mathrm{d} u
	= 2w_k\Phi\Big(
	\frac{(\vtheta-\vmu_k)\trans \mSigma_k^{-1}\valpha_k}{\sqrt{1+\valpha_k\trans \mSigma_k^{-1} \valpha_k}}	\Big)
	\phi\big(\vtheta; \vmu_k, \mSigma_k + \valpha_k\valpha_k\trans\big).	
		\end{equation}
A mixture component in~\eqref{eqn:infiniteMixMain} can also belong to
the normal/independent distribution family introduced in~\cite{lange1993normal}, which includes the multivariate  t-distribution, the slash distribution. In this work, we make use of  the multivariate  t-distribution with $\nu_k$ degrees of freedom as the mixture components, and therefore write  $\alpha_k = \beta_k = \nu_k/2$ so that
	\begin{equation} \label{eqn:phimix:tdist}
	\phi^{(k)}(\vtheta) = w_k \int \phi(\vtheta;\ \vmu_k, v \mSigma_k ) \mathrm{InvGamma}(v; \alpha_k, \beta_k) \mathrm{d} v.
		\end{equation}

\begin{figure}[t!]
	\centering
	\includegraphics[width=0.6\textwidth]{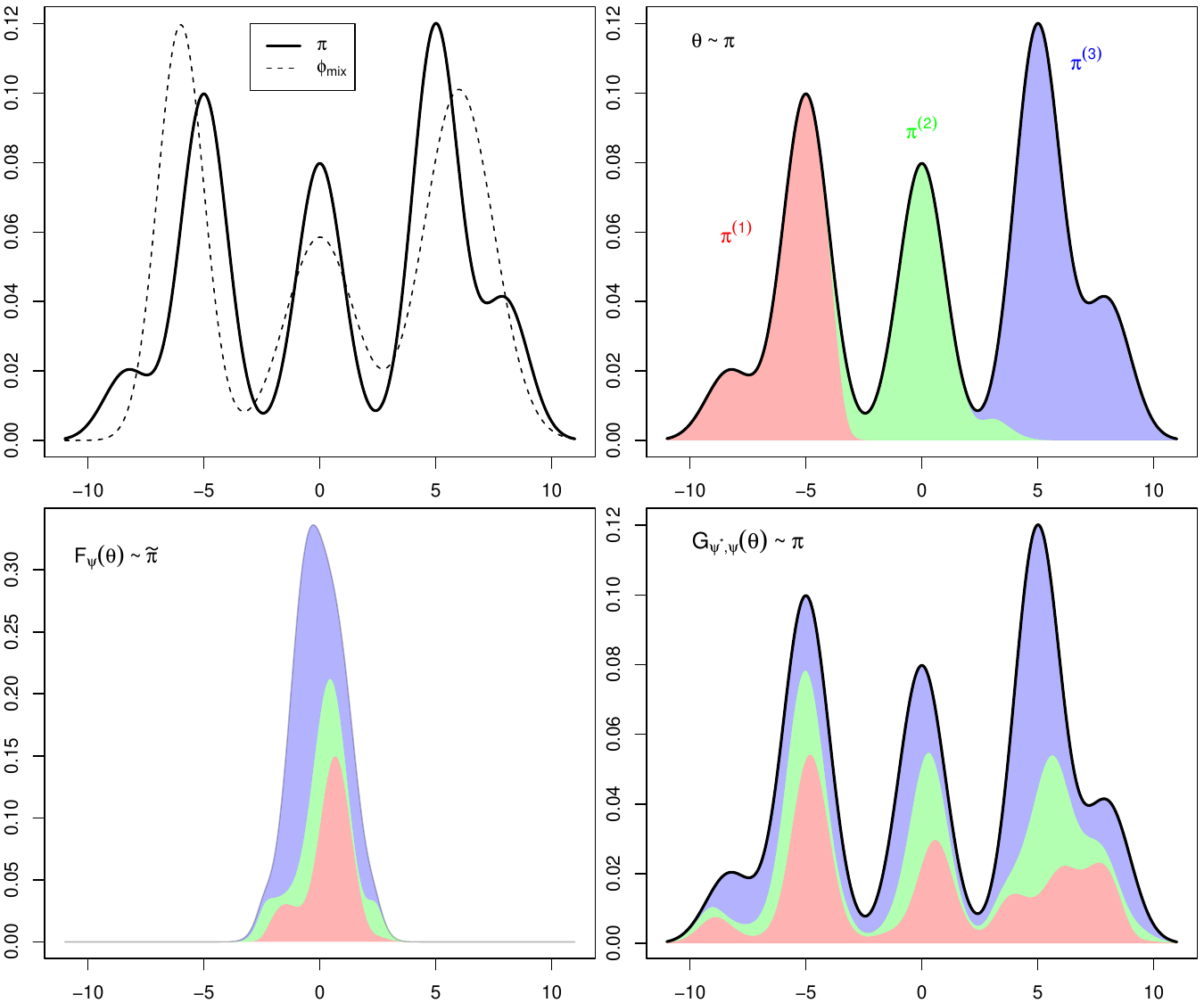} 	
	\caption{\small Top left:  the Gaussian mixture $\phi_{\text{mix}}$ (dashed line) is a rough approximation to the target density $\pi$ (solid line). Top right: the components $\pi^{(1)}$, $\pi^{(2)}$, and $\pi^{(3)}$ of $\pi$ sliced by $\phi_{\text{mix}}$ are shown as red, green and blue  regions, respectively. Bottom left: the  transformed density $\tilde{\pi}$ obtained by applying the stochastic Warp-U   transformation $\sF_\psi$.  Bottom right: applying $\sG_{ \psi',\psi}= \sH_{\psi'}\circ \sF_\psi $ (i.e., the Warp-U transformation and its inverse) exactly recovers the original $\pi$ (black solid line); the shaded colors indicate the proportion of density transported from the original components of $\pi$.}
	\label{fig:warpuexample}
\end{figure}

Define $\zeta = (\psi, u, v)$ as the vector containing the component index $\psi\in\{1,\cdots, K\}$ together with $u(\ge 0)$ and $v$.
A more general form of the Warp-U transformation \citep{wang2022warp}  is the following \textit{distributional normalization}: 
\begin{equation} \label{eqn:warpUtrans:base}
\vtheta^*=\mathcal{F}_{\zeta}(\vtheta) := v^{-1/2}\mS_{\psi}^{-1}(\vtheta-\vmu_{\psi} - u\valpha_{\psi}).
\end{equation}
 When $\vtheta \sim \phi_{\text{mix}}$ in~\eqref{eqn:infiniteMixMain}
and $\zeta = (\psi, u, v)$ are sampled from $	\varpi(\zeta|\vtheta)$:
\begin{equation}
	\varpi(\zeta|\vtheta) = \frac{\phi(\vtheta;\ \vmu_\psi + u \valpha_\psi, v\mSigma_\psi ) p(u,v | \eta_\psi )}{\phi_{\text{mix}}(\vtheta)},\label{eqn:index_conditional}
\end{equation}
we have $\sF_{\zeta}(\vtheta)\sim \mathcal{N}_d(\vzero,\mI)$. That is, the stochastic map $\sF_{\zeta}$  \textit{normalizes} $\phi_{\text{mix}}$ in distribution, since it becomes $\mathcal{N}_d(\vzero,\mI)$.   When $\phi_{\text{mix}}$ approximates $\pi$, it is intuitive that applying the stochastic  transformation
$\sF_{\zeta}$ to $\pi$ (with $\vtheta\sim \pi=q/c$ and $\zeta\sim \varpi(\zeta|\vtheta) $)  also gives  an approximately uni-modal density.  The transformed version of $q$ is given by
\begin{equation}\label{eqn:transfereddensity}
	\tilde{q}(\vtheta^*) =\sum_{k = 1}^{K}w_k
	\int \phi(\vtheta^*)\frac{q(v^{1/2}\mS_k\vtheta^*+\vmu_k + u\valpha_k)}{\phi_{\text{mix}}(v^{1/2}\mS_k\vtheta^*+\vmu_k + u\valpha_k)}
	p(u,v | \eta_k)\,	\mathrm{d} u\, \mathrm{d} v.
\end{equation}
As in \cite{wang2022warp}, the above  $\tilde{q}$ maintains the same normalizing constant as $q$.

To visualize \eqref{eqn:transfereddensity},  consider the case where the employed $\phi_{\rm mix}$ is a Gaussian mixture
$\phi_{\rm mix}^{\rm gauss}$   in~\eqref{eqn:phimix:gauss}, and note in this case $\zeta =\psi$ only contains the component index.  The colored regions in the top right panel of Figure~\ref{fig:warpuexample} display components
$\pi^{(k)}(\vtheta) = \varpi(k|\vtheta) \pi(\vtheta)$  of $\pi$ (black solid curve) induced by $\phi_{\rm mix}^{\rm gauss}$  (the dashed curve in the top left panel). 
 The density of  $\pi^{(k)}$ gets transported towards the origin via the mapping $\sF_\psi$ with $\psi=k$. The bottom left panel of Figure~\ref{fig:warpuexample} is the  Warp-U transformed density $\tilde{\pi}$. The shaded regions represent the proportion of density contributed from the original components of $\pi$ with the corresponding colors.

To estimate the normalizing constant $c$ of $q$,  following  \citet{wang2022warp},  we  can apply~\eqref{calssical_Bridge_estimation}   with densities  $q_1 = \tilde{q}$ in~\eqref{eqn:transfereddensity}  and $q_2 = \phi$.  However, this approach can have high computational cost when evaluating $q$ is expensive . Even when  $\phi_{\rm mix}$ is   a Gaussian mixture  as in \citet{wang2022warp} and integral evaluation in~\eqref{eqn:transfereddensity} is unnecessary, each evaluation of  $\tilde{q}$ requires
$K$  evaluations of $q$, and overall Warp-U bridge estimator requires $K(n_1+n_2)$ evaluations of $q$. 
Thus, although the Warp-U bridge estimator is statistically more efficient than the standard bridge sampling applied to $\target$ and $\phi_{\text{mix}}$, it can be much more expensive.  \citet{wang2022warp} found that, given fixed computational resources, Warp-U bridge sampling is only comparable to standard bridge sampling (and is sometimes slightly worse). 

\subsection{Stochastic Warp-U Bridge Sampling Estimator}
\label{sec:stochasticBridgeEstimation}

To increase computational efficiency, we note that because $\tilde{q}$ in~\eqref{eqn:transfereddensity} preserves the normalizing constant $c$, we can write $c = \sum_{k=1}^K w_k c_k$, where $c_k$ is the normalizing constant of $\tilde{q}_{k}(\vtheta^*)$:
\begin{equation}\label{equationq_2k} \small 
	\tilde{q}_{k}(\vtheta^*)  = \int 	\tilde{q}_{k}(\vtheta^*, u, v) 	p(u,v | \eta_k)	\mathrm{d} u \mathrm{d} v, \text{ with }
	\tilde{q}_{k}(\vtheta^*, u, v) = 	\frac{  \phi(\vtheta^*)  q(v^{1/2}\mS_k\vtheta^*+\vmu_k + u\valpha_k)}{\phi_{\text{mix}}(v^{1/2}\mS_k\vtheta^*+\vmu_k + u\valpha_k)}.
\end{equation}
By the construction of $\vtheta^*=\sF_\zeta(\vtheta)$, we know that $\tilde{q}_k(\vtheta^*)/c_k=p(\vtheta^*|\psi=k)$ is  the conditional density of $\vtheta^*$ given $\psi = k$. Thus the normalized counterpart of \eqref{eqn:transfereddensity} is the  mixture density 
\begin{equation} \label{eqn:warpU:mixturedensity}
	\tilde{\pi}(\vtheta^*) = \frac{\tilde{q}(\vtheta^*) }{c} = \sum_{k=1}^{K} \frac{w_k c_{k}}{c} \frac{\tilde{q}_{k}(\vtheta^*)}{c_{k}} =
	\sum_{k=1}^{K} \tilde{w}_k  \tilde{p}_{k}(\vtheta^*),
\end{equation}
where $\tilde{w}_k=w_k c_{k}/c$ is the  mixture weight for $\tilde{p}_{k}(\vtheta^*)\equiv p(\vtheta^*|\psi=k)$, for all $k$. This immediately suggests that we can apply bridge sampling to estimate each $c_k$ separately, using the draws $\sF_\zeta(\vtheta)$ where the component $\psi$ in $\zeta=(\psi, u, v)$ satisfies $\psi=k$.

\begin{algorithm}[t]
	\caption{Stochastic Warp-U Bridge Sampling Estimator} 		\label{alg:stochasticBridge}	
	\begin{algorithmic}[1]
		{\small \STATE Sample $\vtheta_{1,1},\dots,\vtheta_{1,n_1}\sim \pi$.
			\FOR{ $j=1,\dots,n_1$}
			\STATE (i) sample $\zeta_j = (\psi_{1,j},\, u_{1,j},\, v_{1,j})$ from the probability density $\varpi(\psi, u, v|\vtheta_{1,j})$ in~\eqref{eqn:index_conditional}.  \label{SWB_divide_3}
			\STATE (ii) Set $\vtheta^*_{1,j} = \mathcal{F}_{\zeta_j}({\vtheta}_{1,j}) = v_{1,j}^{-1/2}\mS_{\psi_{1,j}}^{-1}(\vtheta-\vmu_{\psi_{1,j}} - u_{1,j} \valpha_{\psi_{1,j}})$; \label{SWB_divide_4}
			\ENDFOR
			\FOR{ each component $k$ from $1$ to $K$}
			\STATE (A) For $j=1,\ldots, n_2$, sample $\vtheta_{2,k,j} \sim N(0, 1)$, 
			and sample $(u_{2,k,j}, v_{2,k,j})$ from $p(u,v| \eta_k)$; 
			\STATE (B) Apply all $(\vtheta^*_{1,j}, \zeta_j)$ from (i) with $\psi_j=k$ and the  samples from (A) to compute \eqref{rk_estimator}; 
			\ENDFOR
			\STATE Calculate the final estimator $\hat{c}_{\text{SWB}} = \sum_{k=1}^Kw_k\hat{c}_{k}$.}
	\end{algorithmic}
\end{algorithm}

In particular, our estimation of $c_k$ is based on the following bridge identity:
\begin{equation} \label{eqn:bridgenew}
c_k =  \frac{\Expect_{p_2}[\tilde{q}_{k}(\vtheta^*, u, v) \alpha(\vtheta^* ,u,v)]}{\Expect_{p_1}[\phi(\vtheta)\alpha(\vtheta,u,v)]}
\end{equation}
{In the denominator,  the expectation is  with respect to $p_1(\vtheta^*,u,v) = 	\tilde{q}_{k}(\vtheta^*, u, v)  p(u,v|\eta_k)/ c_k$, which is the joint distribution of $(\vtheta^*,u,v)$ given $\psi=k$.
In the numerator, the expectation is taken with respect to the density $p_2(\vtheta^*,u,v) = \phi(\vtheta^*) p(u,v|\eta_k)$, which is an independent coupling of $\phi(\vtheta^*)$ and 
$p(u,v|\eta_k)$.}  Our proposed bridge estimation procedure is detailed in Algorithm~\ref{alg:stochasticBridge}.  {It accepts either i.i.d. samples or MCMC samples post burn-in as input.}
In Lines 3--4,  for each sample $\vtheta_{1,j}$ from $\pi$, it draws an index $\zeta_j = (\psi_{1,j},\, u_{1,j},\, v_{1,j})$ by~\eqref{eqn:index_conditional}, and takes the Warp-U transformation. In Lines 6--9,  
it computes  $c_k$  from  the transformed  samples $\vtheta_{1,j}^*$ for which $\psi_{1,j}=k$. 
For each $k$, we also draw  auxiliary  $\vtheta_{2,k,j}$ from $\phi$ and draw an additional $(u_{2,k,j}, v_{2,k,j})$ from $p(u,v| \eta_k)$. Then, based on~\eqref{eqn:bridgenew}, we evaluate
\begin{equation}\label{rk_estimator}
	\hat{c}_k=\frac{n_{2}^{-1}\sum_{j=1}^{n_{2}}\tilde{q}_{k}(\vtheta_{2,k,j}, u_{2,k,j}, v_{2,k,j}) \times \alpha_k(\vtheta_{2,k,j}, u_{2,k,j}, v_{2,k,j}) }
	{n_{1k}^{-1}\sum_{j\in \sJ_k} \phi(\vtheta^*_{1,j})\times 
		\alpha_k(\vtheta^*_{1,j},  u_{1,j}, v_{1,j})} 
\end{equation}
where	$\sJ_k = \{j:\; \psi_j = k\}$ and $n_{1k} = |\sJ_k|$ is the cardinality  of the index set $\sJ_k$.  
The final estimate of $c$ is the weighted sum $\hat{c}_{\text{SWB}} := \sum_{k=1}^{K}w_k\hat{c}_k$.
We call our estimator $\hat{c}_{\text{SWB}}$ the {\it stochastic} Warp-U bridge (SWB) estimator because it can be viewed as an estimator of the mean normalizing constant of a random unnormalized density.

 Algorithm~\ref{alg:stochasticBridge} requires a total  of $n_1+ Kn_2$ 
evaluations of $q$, much less than that needed for Warp-U bridge estimation. When working with a Gaussian mixture $\phi_{\rm mix}^{\rm gauss}$,   the evaluations of $q$ is $(K-1)n_1$ fewer.  As in  \cite{meng1996simulating}, we can  show that the optimal  $\alpha_k$ for~\eqref{eqn:bridgenew} is  $\alpha_k(\vtheta, u, v) \propto [n_{1k}c_k \phi(\vtheta) +  n_{2k}\tilde{q}_k(\vtheta,u,v)]^{-1} $.
 The theoretical analysis  of~\eqref{rk_estimator}
and the justification of the optimality  of $\alpha_k$ is  in Section~S.4.1 of the online Supplement.

\subsection{A Distribution-Preserving Mass-Swapping  Transport}
\label{sec:distpre}

 Recall that, when $\vtheta\sim \pi$, the  Warp-U transformed random variable ${\vtheta}^*=\mathcal{F}_{\zeta}(\vtheta)$  follows $\tilde{\pi} = \tilde{q}/c$  of ~\eqref{eqn:transfereddensity}.  Our Warp-U sampler then applies the {\it inverse} Warp-U transformation to obtain a draw $\vtheta' = \mathcal{H}_{\zeta'}(\vtheta^*) := \mathcal{F}_{\zeta'}^{-1}(\vtheta^*)$,  where the random index $\zeta'$ is generated to preserve the target distribution $\vtheta' \sim \pi $.  We can achieve this easily and in general by Theorem~\ref{th:preserve}. 


\begin{theorem}
	\label{th:preserve}
	Suppose $\mathcal{F}_{\zeta}(\cdot)$ is bijective for any  given $\zeta$.	Let $\vtheta^*=\mathcal{F}_{\zeta}(\vtheta)$ be a stochastic map from  $\Theta$, the support of a random variable $\vtheta$, to itself, where the random index $\zeta$  has a well-defined joint distribution with $\vtheta$.  Suppose $\zeta' \sim \nu(\zeta|\vtheta^*)$, the conditional distribution of $\zeta$ given $\vtheta^*$.  Then $\vtheta'=\sF^{-1}_{\zeta'}\left(\sF_\zeta(\vtheta)\right)$ and $\vtheta$ are identically distributed.  Furthermore, if $\zeta'$ and $\zeta$ are independent conditional on $\vtheta^*$, then $\vtheta'$ and $\vtheta$ are i.i.d. given $\vtheta^*$. 
\end{theorem}
\begin{proof} Suppose $\zeta, \vtheta$ are drawn from their joint distribution $p(\zeta, \vtheta)$. This  joint distribution determines  the marginal distribution $p(\vtheta)$,  the joint distribution $p^*(\zeta, \vtheta^*)$ of $\zeta$ and  $\vtheta^*=\mathcal{F}_{\zeta}(\vtheta)$, and the conditional distribution $\nu(\zeta|\vtheta^*)$.   Because $\vtheta=\mathcal{F}^{-1}_\zeta(\vtheta^*)$ since $\mathcal{F}_{\zeta}(\cdot)$ is bijective, we see that drawing $\zeta'\sim \nu(\zeta|\vtheta^*)$ is sufficient for $\vtheta'=\mathcal{F}^{-1}_{\zeta'}(\vtheta^*)$ to be identically distributed as $\vtheta$, and that $\vtheta$ and $\vtheta'$ are conditionally independent given $\vtheta^*$ if $\zeta$ and $\zeta'$ are so.
\end{proof}

Consequently, by drawing $\zeta'$ independently from
\begin{equation}  \label{eqn:star_conditional}
	\nu (\zeta|\vtheta^{*}) \propto \varpi(\zeta|\vtheta=\mathcal{H}_{\zeta}(\vtheta^{*})) q (\mathcal{H}_{\zeta}(\vtheta^{*})) \times |\mathcal{H^{'}}_{\zeta}(\vtheta^{*})|,
\end{equation}
where $|\mathcal{H^{'}}_{\zeta'}(\vtheta^{*})|$ is the absolute  value of the Jacobian determinant of $\mathcal{F}_\zeta^{-1}$, we ensure $\sG_{ \zeta',\zeta}= \sH_{\zeta'}\circ \sF_\zeta $ is distribution preserving, and that
its output $\vtheta'$ is conditionally independent of $\vtheta$ given $\vtheta^*$.  This conditional independence, which holds regardless of the distribution of $\vtheta^*$, is critical for preventing our algorithm from being trapped by any particular mode, statistically speaking.  Furthermore,   neither the distribution preserving property nor the conditional independence requires knowledge of how well $\phi_{\rm mix}$ approximates $\pi$.

The bottom right panel of Figure~\ref{fig:warpuexample} illustrates the result of the  two-step stochastic transformation $\sG_{ \zeta',\zeta}= \sH_{\zeta'}\circ \sF_\zeta $  applied to  $\pi$, using the Gaussian mixture distribution $\phi_{\text{mix}}^{\rm gauss}$ in the top left panel. The transformation exactly recovers the original density $\pi$.  More importantly, the  transformation swaps probability masses among the  original components of $\pi$, which are colored in the top right panel. In the bottom right panel of Figure~\ref{fig:warpuexample}, the shaded regions indicate the proportion of density contributed from each original component of $\pi$ with the corresponding color. At most $\vtheta$ locations, there is a considerable amount of density transported from  each of the original components of $\pi$. Consequently, iteratively applying $\mathcal{G}$ to $\pi$ results in the mass between its components mixed rapidly.

Here the transported masses are calculated as follows. Recall, with  the Gaussian mixture distribution $\phi_{\text{mix}}^{\rm gauss}$, the index $\zeta$ only contains the component index $\psi\in\{1,\ldots, K\}$.  Let 
\begin{equation} \label{eqn:contribute}
	\pi_{\zeta', \zeta}(\vtheta) =\pi \big(\sG_{ \zeta',\zeta}^{-1}(\vtheta) \big) \times
	|(\sG_{ \zeta',\zeta}^{-1})'(\vtheta)| \times p(\zeta',\zeta | \sG_{ \zeta',\zeta}^{-1}(\vtheta)),  
\end{equation}
where $|(\sG_{ \zeta',\zeta}^{-1})'(\vtheta)|$ is the Jacobian of $\sG_{ \zeta',\zeta}^{-1}$, and 
$p(\zeta',\zeta |\vxi) = \varpi\big(\zeta |\vxi) \times \nu (\zeta'| \sF_\zeta(\vxi)) $
is the probability of choosing the transformation $\sG_{ \zeta',\zeta}$ at $\vtheta=\vxi$ (with $\vtheta$ being generic notation), that is, the probability of selecting $\zeta$ {\it and} transitioning to $\zeta'$.  Then by applying Theorem~\ref{th:preserve} with discrete index $\zeta$,  we have
$\pi(\vtheta) = 	\sum_{\zeta=1}^K \sum_{\zeta'=1}^K \pi_{\zeta', \zeta}(\vtheta)$,
because Theorem~\ref{th:preserve} tells us that averaging over all the possible transitions must recover $\pi$, since $\vtheta$ and $\vtheta'$ have the same distribution.  Therefore, the function $f^{(\zeta)}(\vtheta) = \sum_{\zeta'=1}^K \pi_{\zeta', \zeta}(\vtheta)$
can be interpreted as the amount of density redistributed or transported (including self-transportation term $\pi_{\zeta, \zeta}(\theta)$) from the $\zeta$-th original component of $\pi$.  The densities $f^{(\zeta)}(\vtheta), \zeta=1, 2, 3$ correspond to the three shaded areas with different colors in the bottom right panel of Figure~\ref{fig:warpuexample}.

\subsection{Warp-U MCMC Sampler}\label{section:sampling}

\begin{algorithm}[t]
	\caption{Warp-U MCMC Sampler 		\label{alg:basicSampler}}
	\hspace*{\algorithmicindent} \small	\textbf{Input:} 
a family of	forward transformations $\sF_{\zeta}$,
	the density $\varpi(\zeta|{\vtheta})$ for selecting  $\sF_{\zeta}$, proposal variance $\sigma^2$, initial value $\vtheta_0$, and the number of samples to be collected $T$.
	
	\begin{algorithmic}[1]
		\FOR{ $t=1,2,\dots,T$}
		\STATE (i) Generate ${\vtheta}^{\text{\tiny{MH}}}$ using a Metropolis-Hasting step with proposal $\mathcal{N}(\vtheta_{t-1},\sigma^2\mI)$.\label{stepMH}
		\STATE (ii) Sample $\zeta$ from 
		$\varpi(\zeta|{\vtheta}^{\text{\tiny{MH}}})$, and set $\vtheta^* = \mathcal{F}_{\zeta}({\vtheta}^{\text{\tiny{MH}}})$.
		\STATE (iii) Sample $\zeta'$ from
		$\nu(\zeta'|\vtheta^{*})$ in~\eqref{eqn:star_conditional}, and set $\vtheta_{t} = \mathcal{H}_{\zeta'}(\vtheta^{*})
		\equiv \mathcal{F}^{-1}_{\zeta'}(\vtheta^{*})$.
		\ENDFOR
	\end{algorithmic}
\end{algorithm}

The redistribution of mass via Warp-U transformations provides a candidate MCMC sampler. Given an  initial sample   $\vtheta_0$,
we can repeatedly apply  the random transformation $\vtheta_{t} = \mathcal{G}_{\zeta',\zeta}(\vtheta_{t-1})$ to generate a sequence of $\vtheta$'s which switches among the target components. However, this switching is insufficient for constructing a valid MCMC sampler, because the resulting Markov chain is not guaranteed to be irreducible. For illustration, consider an example where the auxiliary distribution~\eqref{eqn:infiniteMixMain} is set as $\phi_{\text{mix}}^{\rm gauss}(\vtheta) = \sum_{k=1}^K \phi(\vtheta-\vmu_k) /K$.
In this case, each $\mathcal{G}_{\zeta',\zeta}(\vtheta) = \vtheta  -\vmu_{\zeta} + \vmu_{\zeta'}$ is a shift transformation.  Given any $\vtheta_0$, the sequence $\vtheta_{t}$ can only visit the countable grid inside
$ \Theta = \big\{ \vtheta:\; \vtheta = \vtheta_0 + j_1 \vmu_1 +\cdots+ j_K\vmu_K,\; j_1,\cdots, j_K\in \mathbb{Z}\big\}$,
where $\mathbb{Z}$ is the set of all integers. Hence, when the target $\pi$ is a continuous density over $\bbR^d$, the chain  cannot converge to the target density for any starting value $\vtheta_0$.


Introducing a Metropolis-Hastings (MH) step between consecutive $\mathcal{G}_{\zeta',\zeta}$ transformations resolves this issue, see  Algorithm~\ref{alg:basicSampler}. The algorithm proceeds by executing a random walk MH step at the beginning of each iteration followed by the stochastic transformations $\mathcal{F}_{\zeta}$ and  $\mathcal{F}_{\zeta'}^{-1}$ with random indices $\zeta,\zeta'$.  
The proposed sampler  is still valid if the MH step in Line~2 of Algorithm~\ref{alg:basicSampler} is replaced by other samplers, e.g., the Metropolis adjusted Langevin  or Hamiltonian Monte Carlo samplers. 
 It is worth noting that  Algorithm~\ref{alg:basicSampler} is presented as a general sampler. It can take as input any potentially valid 
family of transformations $\sF_{\zeta}$, and 	a density $\varpi(\zeta|{\vtheta})$ for selecting  $\sF_{\zeta}$. In this work, our primary focus is on the transformation class in~\eqref{eqn:warpUtrans:base}
and the selection density in~\eqref{eqn:index_conditional}. The corresponding implementation details can be found in Section~S.1 of the online Supplement.

As the selection density in~\eqref{eqn:index_conditional} is computed based on a mixture  distribution $\phi_{\text{mix}}$ approximating the target,  the performance of Algorithm~\ref{alg:basicSampler} may depend on the quality of  the approximation.  An adaptive version in Section~S.2 of the online Supplements  performs a periodic update of $\phi_{\text{mix}}$ from the accumulated samples, a strategy particularly effective in low-dimensional settings. Initially,  $\phi_{\rm mix}$ can be configured with a large  $K$  to ensure comprehensive coverage of the state space $\Theta$ for efficient exploration. As more samples are collected,  $\phi_{\rm mix}$ can be gradually refined to improve the quality of sampling.  In high-dimensional settings, a well-chosen initialization of $\phi_{\rm mix}$ is required for effective multi-mode sampling, which often requires knowledge on local modes.  We can use optimization with multiple random initial points to find the local modes, and then initialize  $\phi_{\rm mix}$  using variational inference techniques \citep{lin2019fast,lin2019stein, lin2020handling}. See Section~S.3 for full details.

\subsection{Neural Warp-U Bridge  Sampler and Estimator} \label{sec:neural}
The class of Warp-U transformations in Section~\ref{sec:extendedWarpU} only consists of affine mappings. However, our sampling and estimation methods can be  applied using any invertible mapping and an appropriate density for selecting them, as demonstrated by Theorem~\ref{th:preserve}. This paves the way for developing a wide variety of transformations to improve performance and we illustrate one such extension here. Recall the Warp-U sampler can be constructed with the Gaussian mixture model in~\eqref{eqn:phimix:gauss}, which determines $K$ transformations
$\mathcal{F}_{k}(\vtheta) =  \mS_{k}^{-1}(\vtheta-\vmu_{k})$ and the selection probability 
$\varpi(k|\vtheta) = \phi^{(k)}(\vtheta)/ \phi_{\rm mix}^{\rm gauss}(\vtheta)$.  Instead of these affine transformations,  we consider for each $k$ a mapping $\mathcal{F}_{\mathrm{neu},k}(\vtheta) = \sT_{k}^{-1}
( \mS_{k}^{-1}(\vtheta-\vmu_{k}) )$, which is a composition of the original  $\mS_{k}^{-1}(\vtheta-\vmu_{k})$ and the inverse of a non-affine  mapping $\sT_k$.  The affine  mapping $\mS_{\psi}^{-1}(\vtheta-\vmu_{\psi})$ handles the location and scale adjustment for each mode, while  $\sT_k$ makes additional local refinements.  With the original $\varpi(k|\vtheta) = \phi^{(k)}(\vtheta)/ \phi_{\rm mix}^{\rm gauss}(\vtheta)$ and the new transformation $\mathcal{F}_{\mathrm{neu}, k}(\vtheta)$, the Warp-U transformed target density becomes
\begin{equation} \label{eqn:transfereddensity:neural}
	\tilde{q}_{\rm{neu}}(\vtheta^*) = \sum_{k=1}^{K}  	\tilde{q}_{\rm{neu},k}(\vtheta^*)  =
	\sum_{k = 1}^{K}
	w_k\phi(\sT_k(\vtheta^*))\frac{q(\mS_k 
		\sT_{k}(\vtheta^*)+\vmu_k)}{\phi_{\text{mix}}^{\rm gauss}(\mS_k\sT_{k}(\vtheta^*)+\vmu_k)}\times 
	|\sT_k '(\vtheta^*) |,
\end{equation}
where $ |\sT_k'(\vtheta^*) |$ is Jacobian of  $\sT_k$
at $\vtheta^*$.
The role of $\sT_k$'s can be understood as performing a density transformation 
to each summand in~\eqref{eqn:transfereddensity}, when $\phi_{\rm mix}$ in~\eqref{eqn:transfereddensity}  is the Gaussian mixture. 

A popular approach for density transformation using neural networks is normalizing flows \citep{papamakarios2021normalizing}. In this work, we employ continuous normalizing flows, which are based on ordinary differential equations \citep[Neural ODE;][]{chen2018neural}. Specifically, each nonlinear mapping $\sT_k$ is determined by an ODE of the form: $\mathrm{d} \vtheta^{(t)} = \vv_{\veta_k} (\vtheta^{(t)}  , t)$  for  $t\in[0,1]$. The neural network $ \vv_{\veta_k}$  (parameterized by $\veta_k$) maps 
$(\vtheta^{(t)}  , t)\in\bbR^{d+1}$ to a vector in $\bbR^{d}$. Denote the initial state (at $t=0$)
and final state (at $t=1$) of this ODE as $\vtheta_0$  and $\vtheta_1$, respectively.
These states determine the mapping $\sT_k$ via the relation $\sT_k^{-1}(\vtheta^{(0)}) = \vtheta^{(1)}$ (or equivalently
$\sT_k(\vtheta^{(1)}) = \vtheta^{(0)}$). Furthermore, we can evaluate
$ |\sT_k'(\vtheta^{(1)}) |
=  \exp \big\{\int_0^1 \nabla \cdot \vv_{\veta_k} (\vtheta^{(t)}  , t)\, \mathrm{d} t \big\}.$
In this work,  the neural networks are trained to find each $\sT_k$  that minimizes the KL divergence between the standard Gaussian density $\phi$ and the corresponding  summand
$\tilde{q}_{\rm{neu},k}(\vtheta)$ in~\eqref{eqn:transfereddensity:neural}.  For the detailed training techniques of neural ODE, we refer the reader to \cite{chen2018neural}.

For a given base mixture model $\phi_{\rm mix}^{\rm guass}$,
the above discussion leads to  a family of  transformations $\mathcal{F}_{\mathrm{neu},\psi}(\vtheta) := \sT_{\psi}^{-1}( \mS_{\psi}^{-1}(\vtheta-\vmu_{\psi}) )$, and the selection probability 
$\varpi(k|\vtheta) = \phi^{(k)}(\vtheta)/ \phi_{\rm mix}^{\rm gauss}(\vtheta)$. They can be employed as inputs to the Warp-U sampler in Algorithm~\ref{alg:basicSampler}  for generating samples. 
We refer to this version of  sampler  as the neural Warp-U sampler.  Meanwhile,  based on the transformed target density~\eqref{eqn:transfereddensity:neural}, the Warp-U bridge estimator and the  stochastic Warp-U bridge estimator  can be adapted for normalizing constant estimation of the target~$q$.    These are referred to as neural (stochastic) Warp-U bridge estimators.

\subsection{Theoretical Properties}
\label{sec:theory}

\begin{table}[t]
	\caption{\small Number of evaluations of the unnormalized target density $q$ for different sampling and estimation methods needed to obtain $n_1$ target draws and a total of $n_2$ auxiliary draws which are then used in the estimation step. No. Iters represents the number of iterations (accept and reject), $n_1$ and $\tilde{n}_1$ denote the number of samples and average number of samples at each stage, respectively, {$M$ is the number of stages of the adaptive version of our Warp-U sampler given in supplementary Section~S.2 ($M=1$ for Algorithm \ref{alg:basicSampler}),  $M_s$} is the number of stages of the Generalized Wang-Landau algorithm, $M_l$ is the number of temperature levels in Parallel Tempering. Note that, although the number of target draws ($n_1$) is fixed here, the effective sample size is not.}
	\centering
  \resizebox{0.9\textwidth}{!}{
		\begin{tabular}{r|r|r|r|r|r}
			\toprule
			\multirow{2}{*}{Sampling}  & \multirow{2}{*}{No. Iters}  & \multirow{2}{*}{Sampling Evals.} &\multicolumn{3}{c}{Estimation Evals.} \\
			\cmidrule(r){4-6}
			\ & \ &\ &Bridge &Warp-U & S. Warp-U\\
			\hline
			Warp-U MCMC & $n_1M$ & $Kn_1M$& $n_2$& $Kn_2$& $n_2$    \\
			Generalized Wang-Landau  & $\tilde{n}_1M_s$ & $\tilde{n}_1M_s$& $n_2$&$(K-1)n_1+Kn_2$&$n_2$    \\
			Parallel Tempering  & $2 n_1 M_l$ & $ n_1 M_l$& $n_2$&$(K-1)n_1+Kn_2$&$n_2$    \\
			Unknown Sampling  & $-$ & $-$& $n_1 + n_2$ & $Kn_1+Kn_2$& $n_1+n_2$\\
			\hline
	\end{tabular} }
	\label{tab:ComparisonEvaluation}
\end{table}

Due to space limitation, theoretical properties of the proposed algorithms are provided in Section~S.4 of the online Supplementary Material.  Theorem~S.4 establishes the ergodicity of the proposed Warp-U sampler.  {We identify a set of mild sufficient conditions under which the algorithm, when equipped with a general family of transformations and their selection probabilities, maintains ergodicity. In particular, the sampler is ergodic when  the transformation class in~\eqref{eqn:warpUtrans:base}
and the selection density in~\eqref{eqn:index_conditional} are employed, and when $\phi_{\rm mix}$ is either a mixture of Gaussian, skewed-Gaussian or t-distributions. The neural Warp-U sampler in Section~\ref{sec:neural} is also ergodic when the aboslute values of the Jacobian determinant of $\sF_{\text{neu},k}$'s are bounded away from zero and away from infinity.}

The asymptotic variance of the stochastic Warp-U bridge estimator $\hat{c}_{\text{SWB}}$  in Algorithm~\ref{alg:stochasticBridge}  is derived in Theorem~S.1. 
Additionally, we  theoretically compare $\hat{c}_{\text{SWB}}$ with Warp-U bridge estimator $\hat{c}_{\text{WB}}$ of \cite{wang2022warp}, when the employed $\phi_{\rm mix}$ is a Gaussian mixture.  In this case, Theorem~S.3 shows that $\hat{c}_{\text{SWB}}$
   is  more efficient than $\hat{c}_{\text{WB}}$ in terms of asymptotic variance and  precision per CPU second (Pps). We define PpS  as $1/(\text{RMSE}\times \text{CPU seconds})$, and assume that other computational costs are negligible compared with evaluating $q  = c\pi$. Therefore  CPU seconds are given by $C\cdot E\cdot g(q)$, where $C$ is a constant, $g(q)$ is the time taken to evaluate $q$ once, and $E$ is the number of required evaluations of $q$. The comparative values of $E$ are shown in the right three columns of Table \ref{tab:ComparisonEvaluation}, where $n_1$ and  $n_2$ denote the number of the samples from the target and auxiliary distribution, respectively. The auxiliary distribution is the standard Gaussian $\phi$ for Warp-U bridge estimation and stochastic bridge estimation, and $\phi_{\text{mix}}$ for classical bridge sampling. The number of Warp-U bridge estimation target evaluations is lower in the case of Warp-U MCMC sampling because some of the necessary evaluations have   been computed during the sampling stage. 
   
\section{Simulation Studies}\label{sec:simulationstudies}
We present three simulation studies that demonstrate both the effectiveness and the limitations of our proposals and their comparands.
Setting I compares the Warp-U samplers to parallel tempering (PT), particularly in high-dimensional settings. Setting II and III examines the variants of Warp-U bridge estimator combined with various sampling strategies. The detailed simulation setup and additional results  can be found in  Section~S.7 of the online Supplement. Section~S.9 presents simulation studies that demonstrate the effectiveness of the adaptive version of our Warp-U sampler presented in Algorithm~S.1 in low-dimensional settings, and our initial study of challenges in the high-dimensional settings.

\noindent \textbf{Setting I: Comparison of  Samplers}.
\cite{woodard2009sufficient} showed that  PT converges slowly in certain high-dimensional contexts, especially when the target distribution is a mixture of Gaussians with distinct variances. We examine the performance of PT and Warp-U sampler in a similar high-dimensional  setting. 
Here, the   target density  is a  mixture  of two skewed t-distributions \citep{gupta2003multivariate}, whose mean vectors, skewness vectors and covariance matrices are randomly generated. 

For our Warp-U sampler,  Step~(i) of Algorithm~\ref{alg:basicSampler} is executed with a Hamiltonian  Monte Carlo (HMC) step.  In Step~(ii), the forward transformations in~\eqref{eqn:warpUtrans:base} are selected by the density in~\eqref{eqn:index_conditional}.  We consider three auxiliary distributions $\phi_{\rm mix}$ in~\eqref{eqn:index_conditional}: 
$ \phi^{\rm gauss}_{\rm mix}$ with two Gaussian components in~\eqref{eqn:phimix:gauss}, 
$\phi^{\rm skew}_{\rm mix}$ with two skewed-Gaussian components in~\eqref{eqn:phimix:skew},
$\phi^{t}_{\rm mix}$ with two t-distribution components in~\eqref{eqn:phimix:tdist}.
Different choices of $\phi_{\rm mix}$ lead to three specific versions of the Warp-U sampler, denoted as
WarpU($\phi^{\rm gauss}_{\rm mix}$),
WarpU($\phi^{\rm skew}_{\rm mix}$) and WarpU($\phi^{t}_{\rm mix}$), respectively.
These $\phi_{\rm mix}$'s are initialized to fit  the target distribution via variational inference.

We also compare four versions of parallel tempering samplers.
The first version, \textit{PT-V}, run the PT chains on the annealed target path $q_t = q^t $  (for $t>0$). The second version, \textit{PT-G}, runs the PT chains on the geometric path: $q_t = q^t\times [\phi_{\rm mix}^{\rm gauss}]^{1-t}$ for $t\in[0,1]$. The chain with $t=0$ is run by drawing independent samples from $q_0=\phi_{\rm mix}^{\rm gauss}$, and for this reason the PT algorithms have the same information about the target mode locations as our Warp-U sampler.
For both PT-V and PT-G, the chains with $t>0$ execute HMC steps for update. The third version,  \textit{PT-V-WarpU($ \phi^{\rm gauss}_{\rm mix}$)}, is the same as PT-V, except that each PT chain executes one step of the WarpU($\phi^{\rm gauss}_{\rm mix}$) sampler for an update.  The forth version, \textit{PT-G-WarpU($ \phi^{\rm gauss}_{\rm mix}$)}, is the same as PT-G, except that the chains with $t>0$ execute the WarpU($\phi^{\rm gauss}_{\rm mix}$) sampler for updates.  In other words, the last two versions run multiple WarpU($\phi^{\rm gauss}_{\rm mix}$) samplers in parallel, with $q_t$'s as their targets.  {The temperature grid needed for PT was  chosen via the algorithm proposed in \cite{atchade2011towards}.}

\begin{figure}
	\centering
	\includegraphics[width=0.3\textwidth]{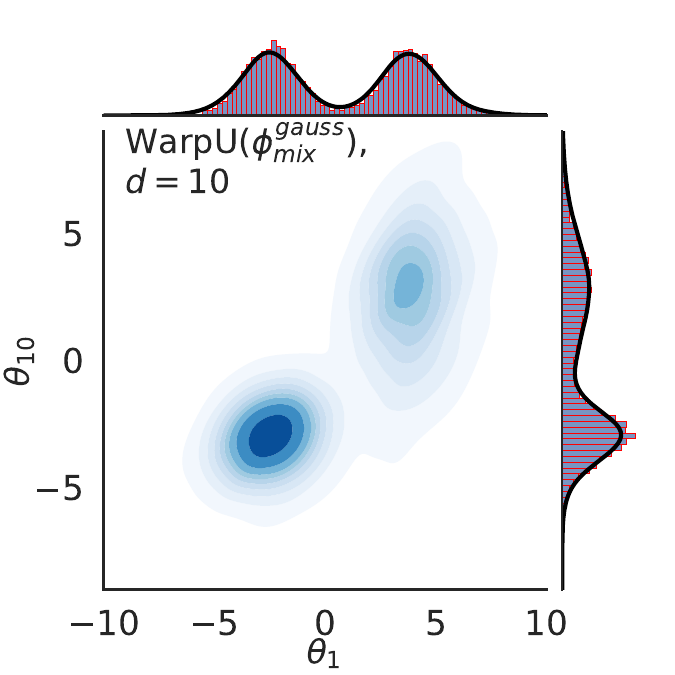}
	\includegraphics[width=0.3\textwidth]{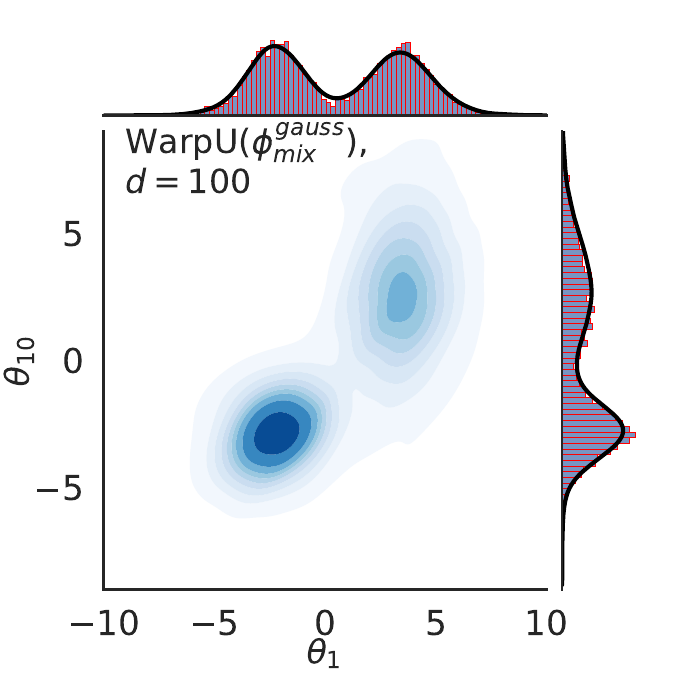}
	\includegraphics[width=0.3\textwidth]{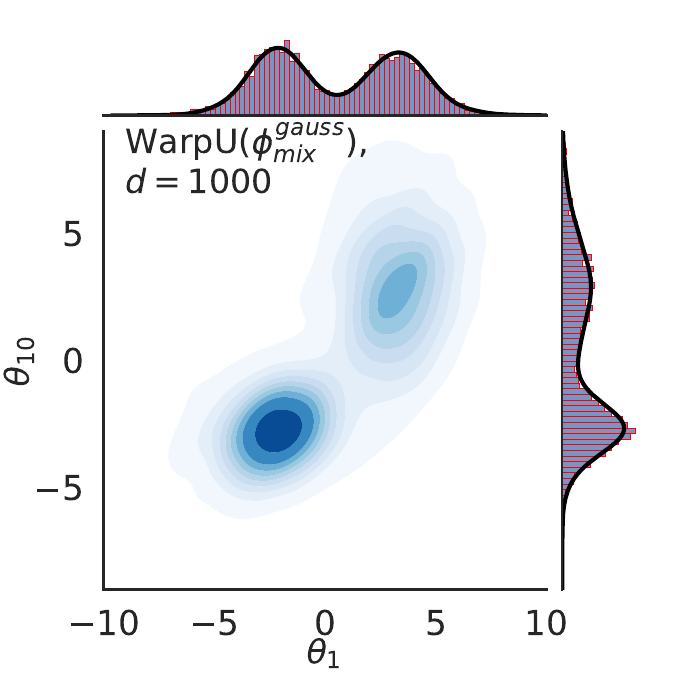}
	\includegraphics[width=0.3\textwidth]{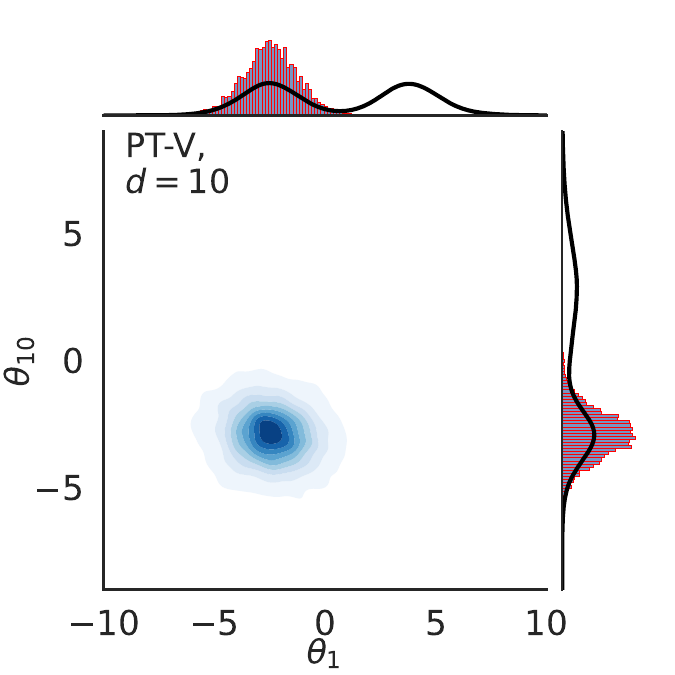}
	\includegraphics[width=0.3\textwidth]{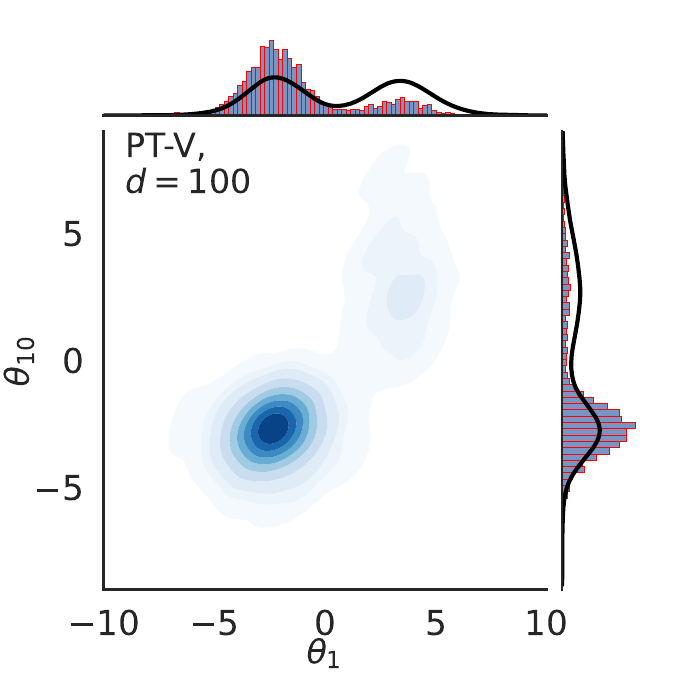}
	\includegraphics[width=0.3\textwidth]{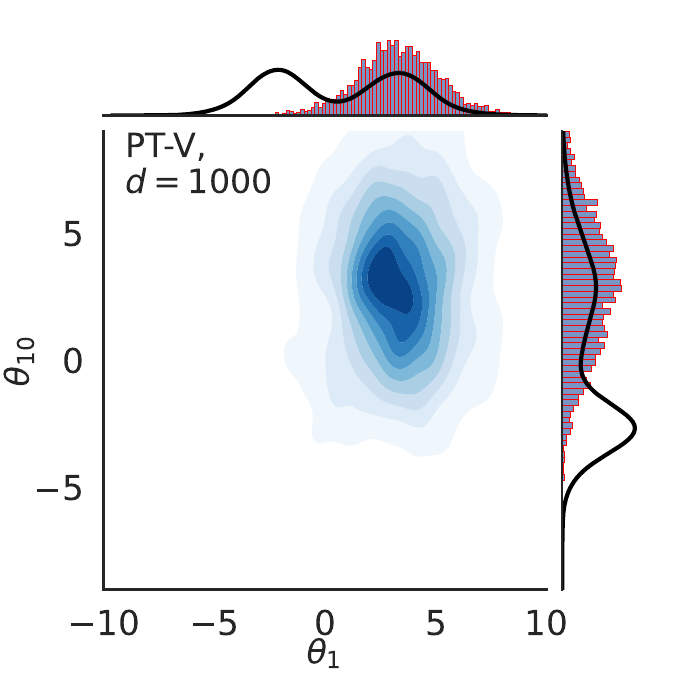}
	\includegraphics[width=0.3\textwidth]{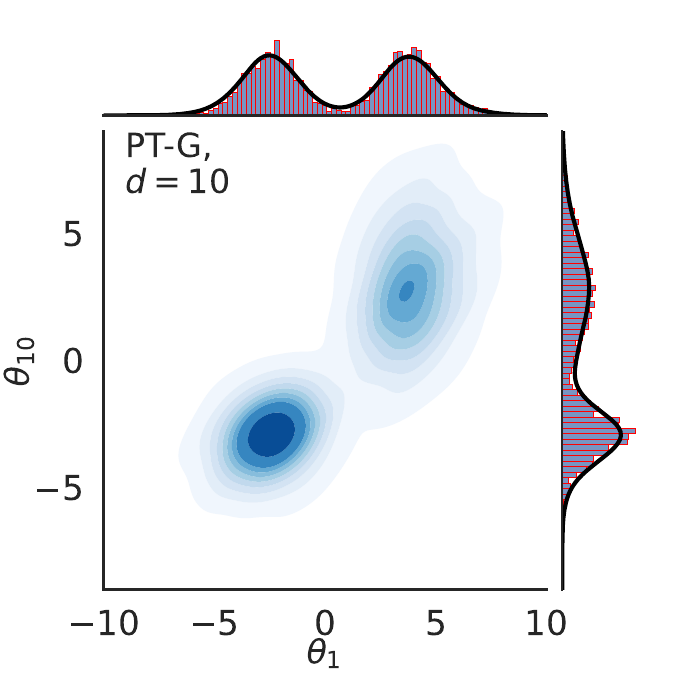}
	\includegraphics[width=0.3\textwidth]{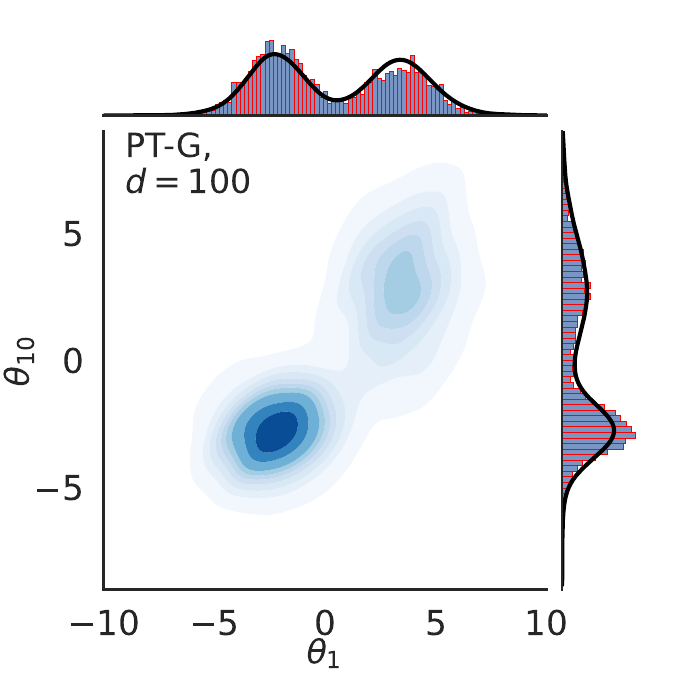}
	\includegraphics[width=0.3\textwidth]{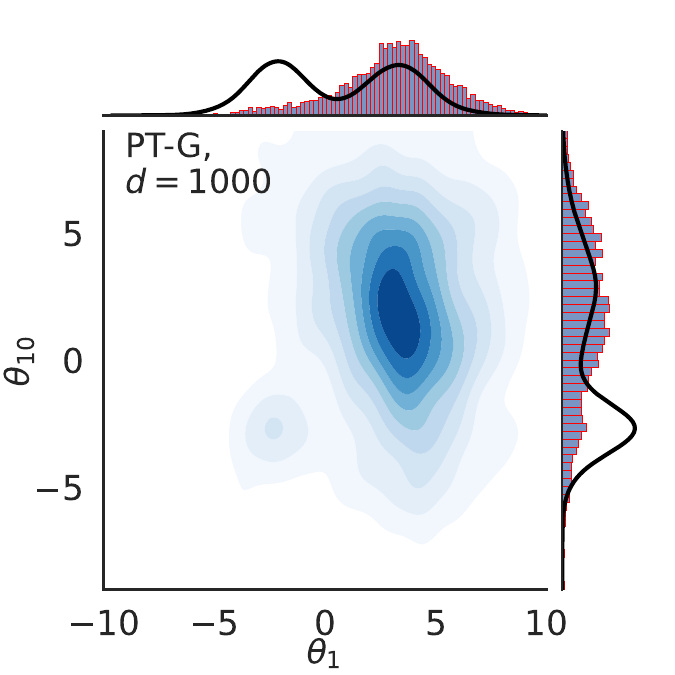}
	\includegraphics[width=0.3\textwidth]{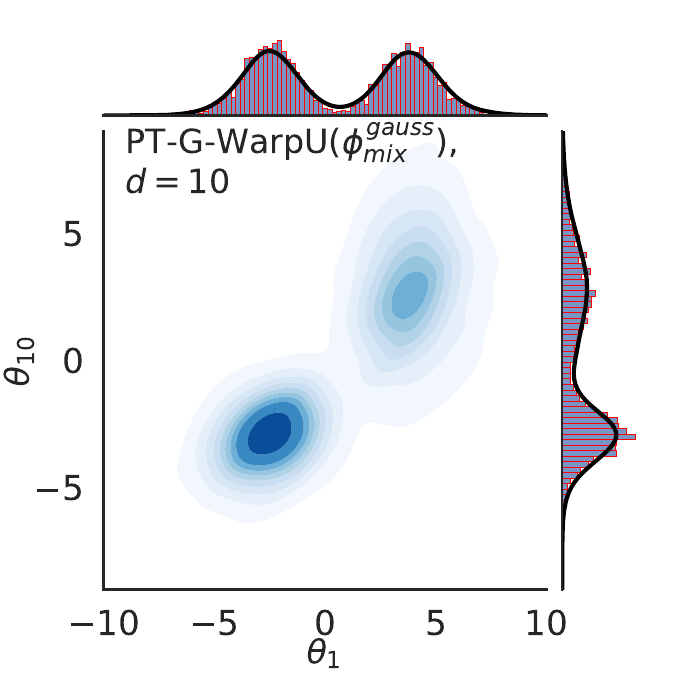}
	\includegraphics[width=0.3\textwidth]{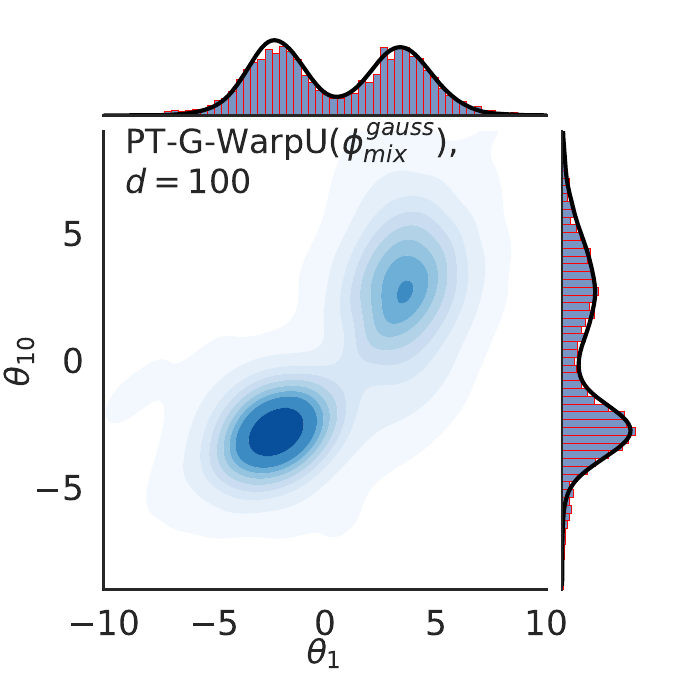}
	\includegraphics[width=0.3\textwidth]{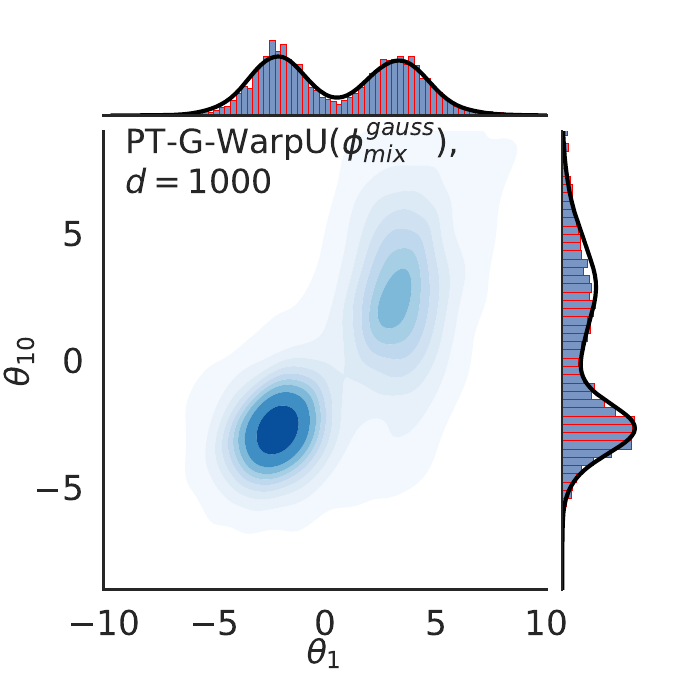}
	\caption{\small Four rows correspond to samplers WarpU($ \phi^{\rm gauss}_{\rm mix}$), PT-V, PT-G, and PT-G-WarpU($ \phi^{\rm gauss}_{\rm mix}$), respectively.	The three columns correspond to dimension $d=10$, $100$, and $1000$, respectively. Each panel shows the density plot for $(\theta_1, \theta_{10})$, which is drawn by one run of  a MCMC sampler. Each panel also shows the marginal histogram  for the Markov chain samples, and the black curves are the kernel density estimates for i.i.d. samples.	 \label{fig:highpt:simu1:new}
	}
\end{figure}

\begin{figure}[t]
	\centering
	\includegraphics[width=0.9\textwidth]{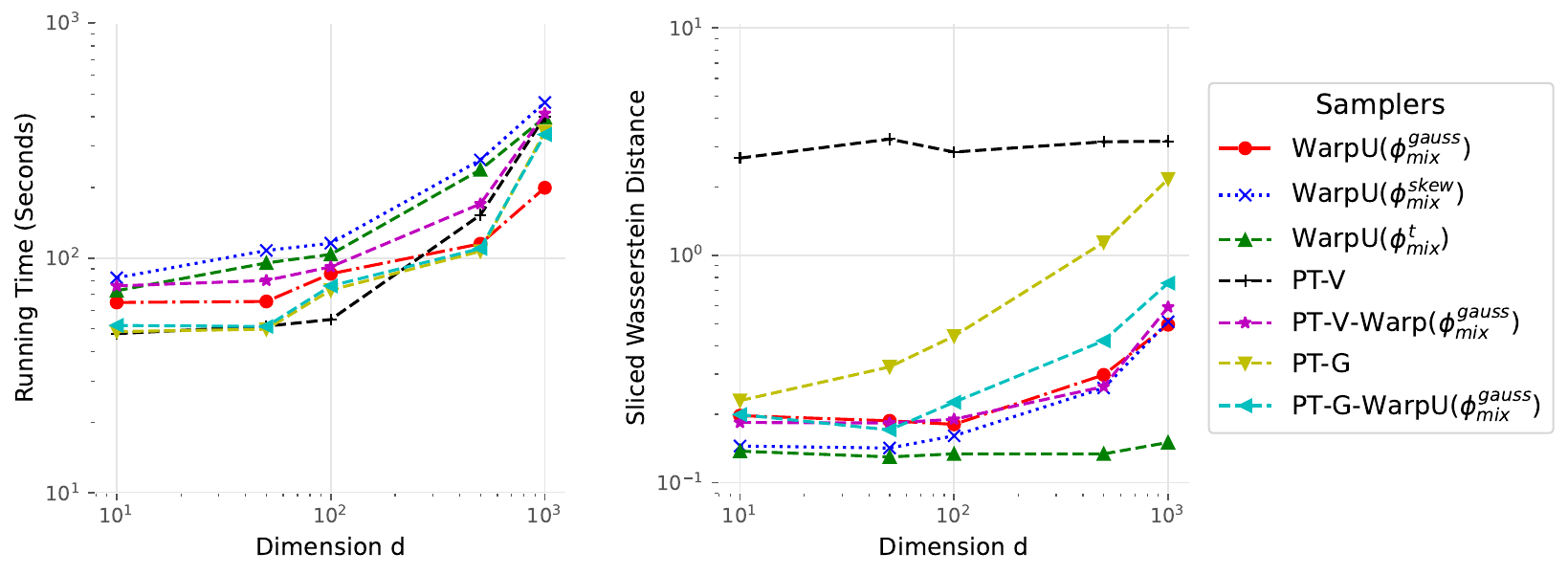}
	\caption{\small Left: average running time for different samplers across dimensions. Right: 
 the sliced Wasserstain distance between MCMC samples and target i.i.d samples as dimension $d$ grows. The distance is computed for the joint marginal $(\theta_1,\theta_d)$ and is averaged across 40 simulation replicates. \label{fig:wass:simu1:new} }
\end{figure}

For each dimension \(d\in\{10,50,100,500, 1000\}\), we ran each algorithm 40 times. In each run, the initial sample was drawn from a standard Gaussian distribution, with the first 100 samples discarded and the subsequent 6000 samples retained. The results of one randomly selected simulation are shown in Figure~\ref{fig:highpt:simu1:new}. The four rows correspond to WarpU($ \phi^{\rm gauss}_{\rm mix}$), PT-V, PT-G and PT-G-WarpU($ \phi^{\rm gauss}_{\rm mix}$), respectively.  From the figure, we can see WarpU($ \phi^{\rm gauss}_{\rm mix}$) effectively captures the two modes across dimensions, while PT-V  struggles with mode jumping even for \(d = 10\). With mode information, PT-G  performs well in low to moderate dimensions. However, for \(d = 1000\), the samples from PT-G exhibit poor mixing.  The fourth row demonstrates that combining PT-G with the Warp-U sampler enhances sample quality for  PT-G. Nevertheless, it still underperforms compared to  WarpU($ \phi^{\rm gauss}_{\rm mix}$).

The average running time of the samplers are in the left panel of  Figure~\ref{fig:wass:simu1:new}, all on log scale. 
Samplers WarpU($\phi^{\rm skew}_{\rm mix}$) and WarpU($\phi^{t}_{\rm mix}$) are most costly, due to the sampling of the skewness and scale parameters $(u,v)$.  As the dimension grows, PT-G becomes more costly, as the target distributions (and their gradients) along the geometric path is more difficult to evaluate.  In the right panel of Figure~\ref{fig:wass:simu1:new}, we show the sliced Wasserstain distance between the MCMC samples and the i.i.d samples from  the target density, computed for two dimensions  $(\theta_1, \theta_{10})$ and averaged across 40 simulation replicates.    The Warp-U sampler with $\phi_{\text{mix}}^t$  consistently generates high-quality samples.  Warp-U sampler with  $\phi_{\text{mix}}^{\rm skew}$ is the second best. Warp-U  with \(\phi^{\rm gauss}_{\rm mix}\) and PT-V-WarpU($\phi^{\rm gauss}_{\rm mix}$) have similar performance. In contrast, PT-V and PT-G have the lowest and second-lowest quality, respectively.


\begin{figure}[t]
	\centering
	\includegraphics[width=0.3\textwidth]{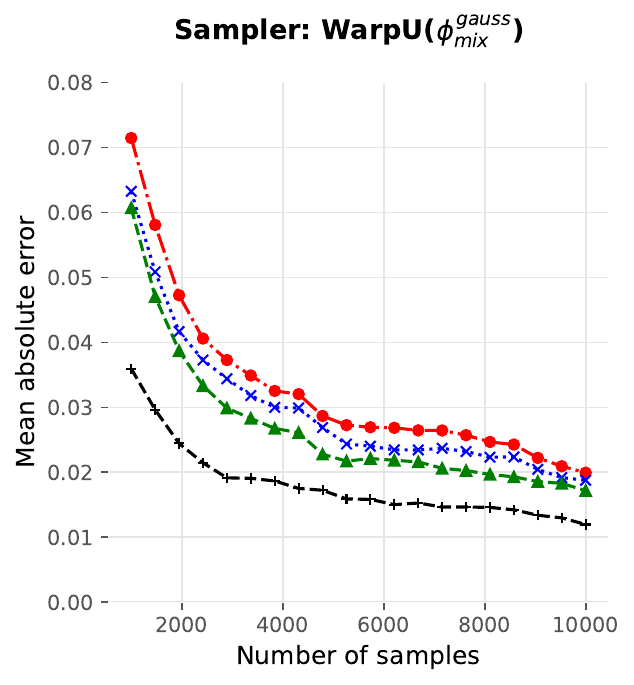} 
	\includegraphics[width=0.3\textwidth]{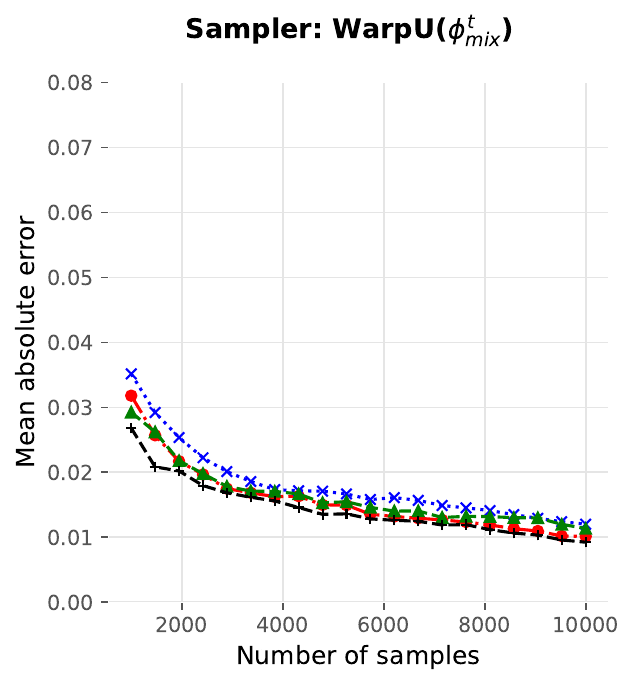} 
	\includegraphics[width=0.3\textwidth]{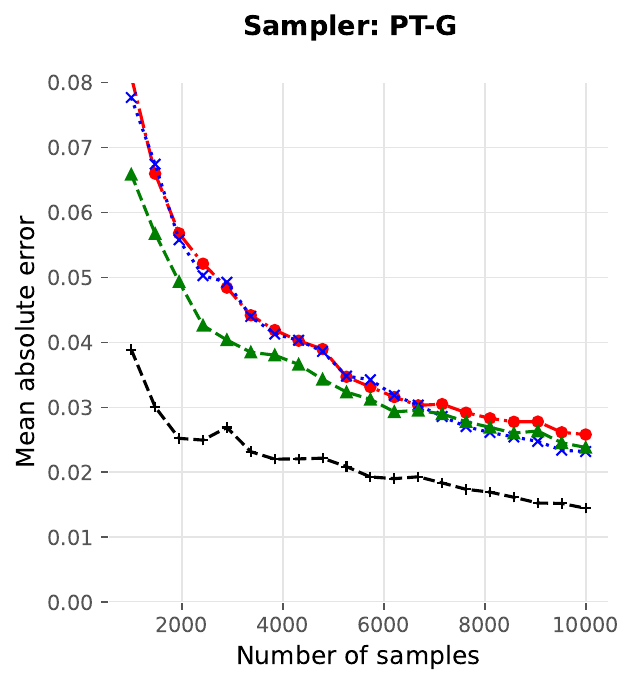} 
	\includegraphics[width=0.3\textwidth]{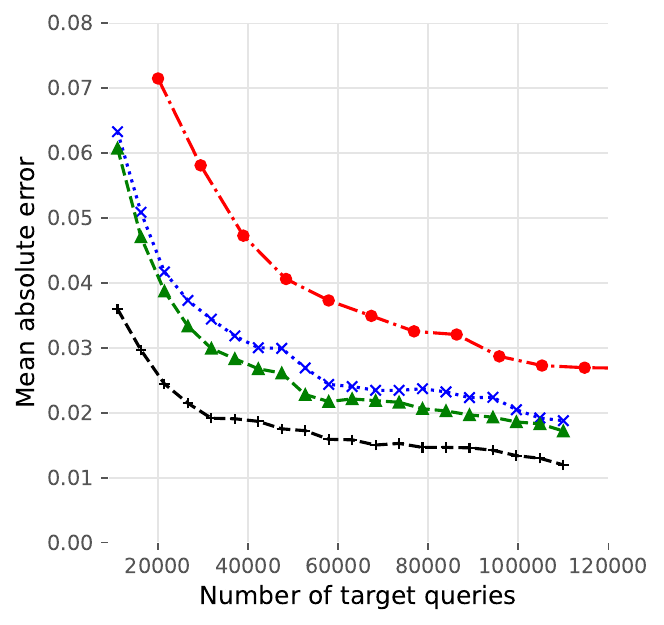} 
	\includegraphics[width=0.3\textwidth]{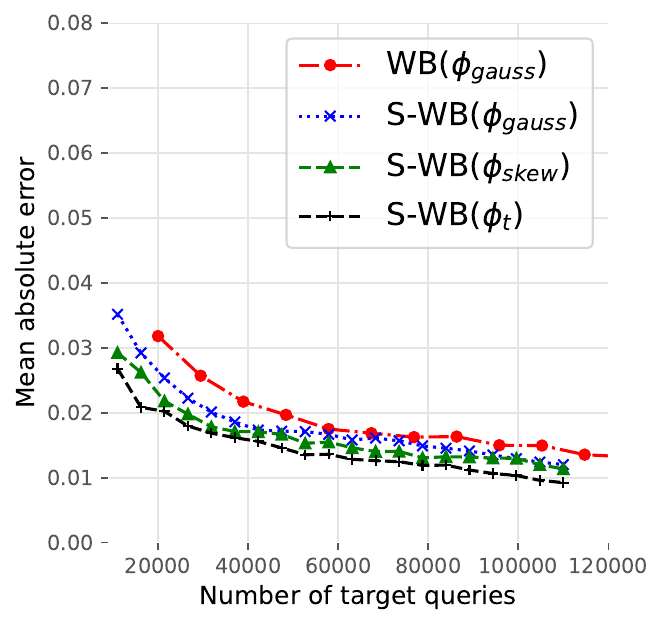} 
	\includegraphics[width=0.3\textwidth]{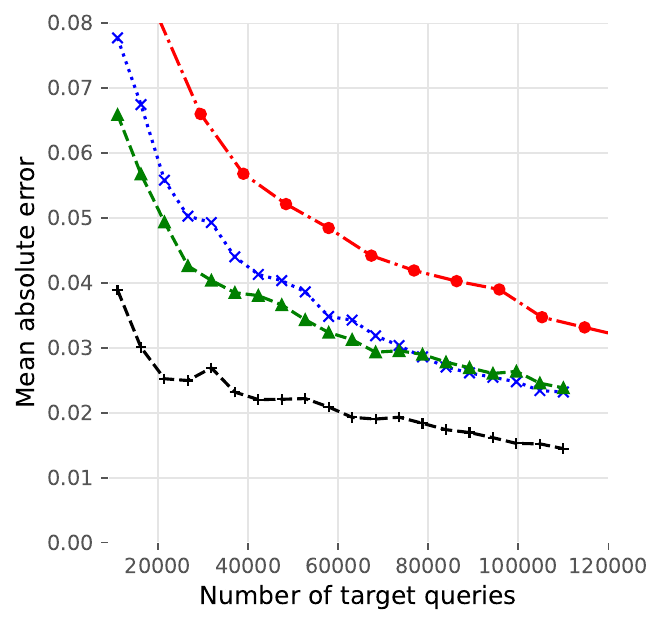} 
	\caption{\small Results for the second simulation setting. Mean absolute errors for the normalizing constant estimators, either as a function of the number of input samples (the first row), or as a function of the number of target evaluations (the second row). The three columns correspond to samples from  WarpU($\phi^{\rm gauss}_{\rm mix}$), WarpU($\phi^{\rm t}_{\rm mix}$)  and PT-G, respectively. 		}
	\label{fig:samplerestimator1}
\end{figure}

\begin{figure}[t]
	\centering
	\includegraphics[width=0.93\textwidth]{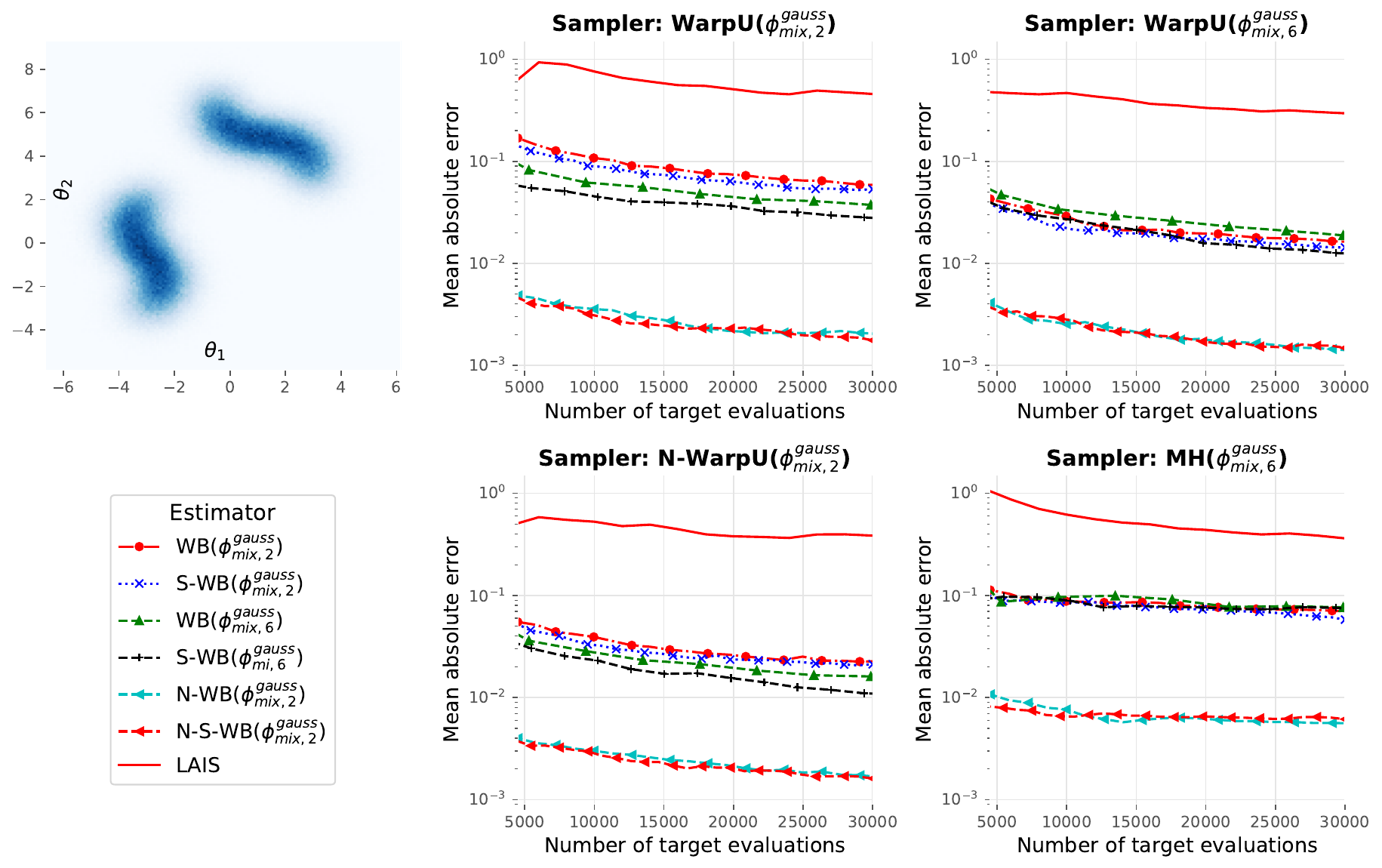} 
	\caption{\small Results for the third simulation setting.  Top left: the density plot for the first two dimensions of the S-shaped target distribution.   The four panels in the middle and right columns correspond to samples from  WarpU($\phi^{\rm gauss}_{\rm mix,2}$), WarpU($\phi^{\rm gauss}_{\rm mix,6}$),  N-WarpU($\phi^{\rm neu}_{\rm mix,2}$), and  MH($\phi^{\rm neu}_{\rm mix,6}$), respectively. 	The mean absolute errors for the normalizing constant estimators are shown as a function of the number of target evaluations.	}
	\label{fig:samplerestimator2}
\end{figure}

\noindent \textbf{Setting II: Comparision of Combined Samplers and Estimators}.
 In this setting, the target density  is a $30$-dimensional mixture of $10$  skew-$t$ distributions with degree of freedom $10$.   S-WB is performed with one of the following auxiliary  distributions with 10  components: the mixture of Gaussian $\phi^{\rm gauss}_{\rm mix}$, the mixture of skewed-Gaussian
$\phi^{\rm skew}_{\rm mix}$ and the mixture of t-distribution $\phi^{\rm t}_{\rm mix}$ with 20 degree of freedom. Meanwhile, WB is only computed with $\phi^{\rm gauss}_{\rm mix}$. 
 We investigate their performance of the estimators {\it given}   $10,000$ samples for some specific sampler described in Setting I, discarding the first 500 samples.

 Figure~\ref{fig:samplerestimator1}  shows the mean  absolute   error of the estimators, either as versus the number of input samples (the first row), or versus the number of target evaluations (the second row), averaging over 100 replicates. The target evaluations are counted  in the estimation stage {\it given} the samples.
 The three columns correspond to samples from  WarpU($\phi^{\rm gauss}_{\rm mix}$), WarpU($\phi^{\rm t}_{\rm mix}$)  and PT-G, respectively. Note S-WB with either auxiliary distribution consistently performs as well or better than WB, both in terms of sample efficiency and computational efficiency (i.e., less target queries given the Monte Carlo samples). In particular,  S-WB  with $\phi^{\rm t}_{\rm mix}$ has the best performance, regardless of the sampling method used. 
 
 In terms of sample quality,  WarpU($\phi^{\rm t}_{\rm mix}$)  sampler  in the middle column is  the best. 
Compared with PT-G, the samples from   WarpU($\phi^{\rm gauss}_{\rm mix}$)  leads to better estimate when the sample size is larger than 2000. This indicates PT-G has better sample quality during the initial phase of the chain, but WarpU($\phi^{\rm gauss}_{\rm mix}$) soon outperforms it as the chain runs longer. 

\noindent \textbf{Setting III: Comparison of combined samplers and estimators}.
In the third setting, the target density  is a $30$-dimensional mixture of two S-shaped distributions, as shown in the top-left panel of Figure~\ref{fig:samplerestimator2}.  
For the Warp-U bridge estimators, we consider two auxiliary distributions  $\phi^{\rm gauss}_{\rm mix,2}$ and  $\phi^{\rm gauss}_{\rm mix,6}$, which are Gaussian mixtures with $K=2$ and $K=6$ components, respectively. They were fitted to the target via variational inference.  To generate samples,  we consider two Warp-U samplers: WarpU($\phi^{\rm gauss}_{\rm mix,2}$),  WarpU($\phi^{\rm gauss}_{\rm mix,6}$). We also implement a Metropolis-Hasting sampler
with independent proposals drawn from  $\phi^{\rm gauss}_{\rm mix,6}$, denoted as MH($\phi^{\rm gauss}_{\rm mix,6}$). Meanwhile, we consider the neural Warp-U bridge  estimators (N-WB($\phi^{\rm gauss}_{\rm mix,6}$)), its stochastic counterpart (N-S-WB($\phi^{\rm gauss}_{\rm mix,6}$)),
  and the neural sampler N-WarpU($\phi^{\rm gauss}_{\rm mix,2}$).
They all  adopt $\phi^{\rm gauss}_{\rm mix,2}$ as their base distribution as discussed in Section~\ref{sec:neural}.

{The layer adaptive importance sampling method  \citep[LAIS,][]{martino2017layered}  is also compared as an estimator.  LAIS has two layers. For the Monte Carlo layer, we use our Warp-U sampler to generate samples to run 100 parallel chains, as our sampler has been demonstrated to be  effective for multi-modal sampling.  In the importance sampling (IS) layer,  additional samples  from an auxiliary mixture model are used to estimate the normalizing constant. Because we compare methods in their estimation stages, we count the number of target distribution queries in the IS layer for LAIS. }

 Figure~\ref{fig:samplerestimator2} presents MAE for the normalizing constant estimators as a function of the number of target evaluations.   The neural (stochastic) Warp-U bridge estimators outperform the other estimators significantly, achieving approximately an order-of-magnitude improvement. 
 Moreover, stochastic Warp-U bridge estimators  generally performs better than the traditional Warp-U bridge estimators. The results in Figure~\ref{fig:samplerestimator2} also  indicate the bridge estimators are more accurate than the importance sampling estimator in this case.  Regarding sample quality, the sampler WarpU($\phi^{\rm gauss}_{\rm mix,6}$) is the best, and N-WarpU($\phi^{\rm gauss}_{\rm mix,2}$) is comparable. In contrast, the samples generated by  MH($\phi^{\rm gauss}_{\rm mix,6}$)
result in high error estimates.

More simulation studies are included in Section~S.9. Sections~S.9.1
 and~S.9.2  contain studies of our adaptive sampler for mode discovery in low dimensional settings, where 
 the target distribution is a 4-dimensional Gaussian mixture with 5 components, and  a 10-dimensional mixture of 25 multivariate skewed t-distributions, respectively. We also made comparisons with the Generalized Wang-Landau  algorithm \citep{liang2005generalized}. Sections~~S.9.3 and~~S.9.4 document our initial efforts to tackle high-dimensional challenges. Since then, substantial progress has been made, as reported in this revision. 

\section{Exoplanet Detection Using Radial Velocity Data}\label{realdataanalysis}

In astronomy, one of the most successful approaches for detecting exoplanets is the radial velocity (RV) method (exoplanets are planets outside our Solar System). The radial velocity of a star is its velocity towards or away from the Earth in meters per second (m/s). When an exoplanet orbits a star, the gravitational force of the planet impacts the RV of the star, and RV data can therefore be used to detect exoplanets. Consider a candidate model $\mathcal{M}$ for capturing the  physical system and noise, e.g., $\mathcal{M}$ might be a Keplerian model for a single exoplanet orbiting a star with Gaussian measurement noise. From a Bayesian perspective, it is natural to compute the Bayesian evidence of the model $\mathcal{M}$:
\begin{equation}\label{bayesianevidence}
	\mathcal{Z} \equiv p(\vd |\mathcal{M}) = \int  p(\vd | \vtheta,\mathcal{M})p(\vtheta|\mathcal{M})\mud\vtheta,
\end{equation}
where $\vd$ is the RV data and $\vtheta$ denotes the parameters of  $\mathcal{M}$. Here $p(\vd | \vtheta,\mathcal{M})$ is the likelihood function and $p(\vtheta|\mathcal{M})$ as the prior. The Bayesian evidence $\mathcal{Z}$ is the normalizing constant of the posterior distribution of $\vd$  given up to proportionality by $\propto  p(\vd | \vtheta,\mathcal{M})p(\vtheta|\mathcal{M})$, and represents the evidence in support of the model $\mathcal{M}$. 

Consider the following model for the observed RV $v_i$ at time $t_i$:
$	v_i = v_{\text{pred}}(t_i|\vtheta) + \epsilon_i$, 
where $v_{\text{pred}}$ is a Keplerian model for the planetary system \citep[see][]{danby1988fundamentals, loredo2012bayesian},   $\vtheta$ denotes the physical parameters, and  $\epsilon_i$ is a noise term; see Section~S.8 for details. One important statistical feature of RV data is that the noise exhibits correlation across observations. In accordance with \citet{rajpaul2015gaussian} and \citet{jones2022improving}, we assume $\vepsilon=(\epsilon_1,\ldots,\epsilon_n)\sim \mathcal{N}(\mathbf{0}, \mSigma)$.
We model the covariance matrix $\mSigma$ by
\begin{equation}
	\mSigma_{ij} = \kappa_{ij} + \delta_{ij}(\sigma_i^2 + \sigma_L^2), \label{model2}
\end{equation}
where $\kappa_{ij}$ is a quasi-periodic kernel, $\delta_{ij}$ is the Kronecker delta, $\sigma_i^2$ is the variance due to measurement error, and $\sigma_L^2$ captures additional variation.   The quasi-periodic kernel is 
\begin{equation}
	\kappa_{ij} = \alpha^2 \exp\left[-\frac{1}{2}\left\{\frac{\sin [\pi (t_i - t_j)/\tau]}{\lambda_p^2} + \frac{t_i - t_j}{\lambda_e^2} \right\} \right],\label{model3}
\end{equation}
and the kernel hyperparameters treated as known in our dataset (described below) are fixed at $\alpha = \sqrt{3} \text{ meters}/\text{second}$, $\lambda_e = 50.0$ days, $\lambda_p = 0.5$ (unit-less), and $\tau = 20$ (days).

\begin{figure}[t]
	\centering
	\includegraphics[width=0.8\textwidth]{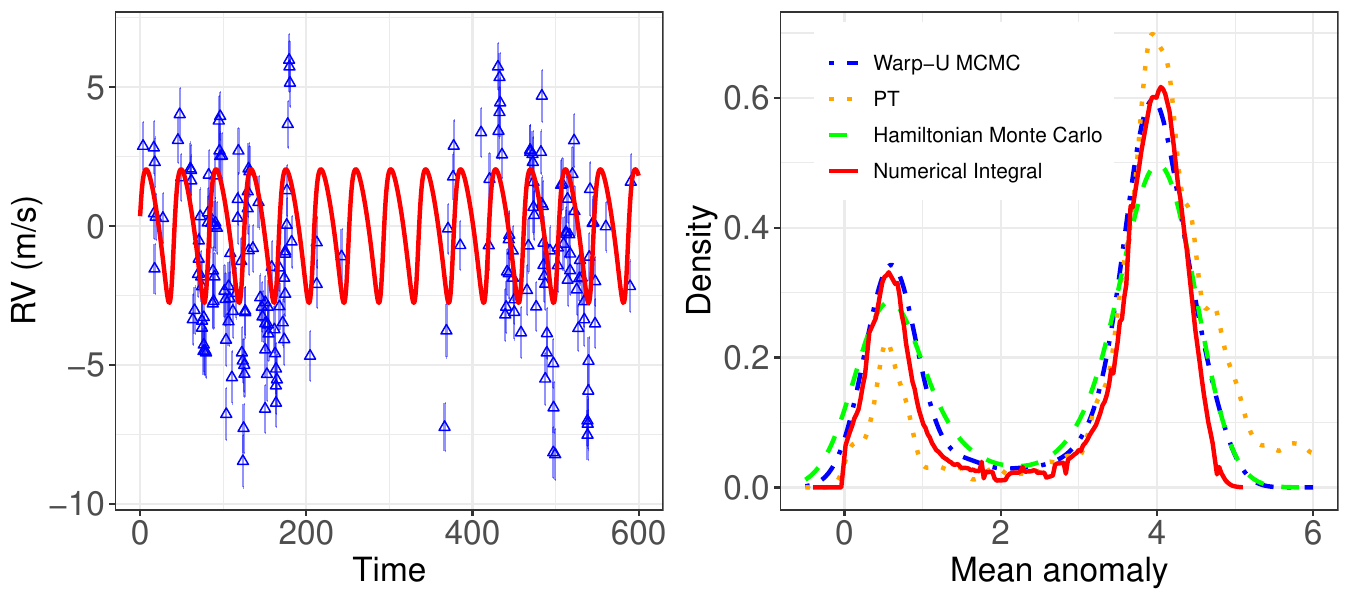} 
	\centering
	\caption{\small The left panel shows the true radial velocity signal as a function of time (solid red line), and observed measurements and their  associated measurement errors $\sigma_i^2$ (blue triangles and vertical bars). Note that the measurement errors represent errors typically reported by an observing telescope, and as can be seen in (\ref{model2}) and the left panel they do not account for all uncertainties. The right panel shows the marginal posterior distribution of the mean anomaly parameter. It compares the estimated densities using the samples obtained by Warp-U MCMC (dash-dot blue line), PT (dotted orange line) and HMC (dashed green line). The solid red line is the estimated target density by the numerical integral. }
	\label{fig:realdatasampling}
\end{figure}

Our dataset, plotted in the left panel of Figure \ref{fig:realdatasampling},  consists of $n=200$ simulated observations from the Extremely Precise Radial Velocities (EPRV3) Evidence Challenge which compared Bayesian evidence estimates produced by different statistical  methods in the context of RV exoplanet detection \citep{nelson2018quantifying}. Each observation comes with the time of measurement and its measurement error in terms of standard deviation.

\noindent\textbf{Comparison of Samplers}. We compare the performance of our adaptive Warp-U MCMC sampler  to that of PT and Hamiltonian Monte Carlo (HMC) \citep[see][]{betancourt2015hamiltonian,neal2011mcmc}, in terms of its ability to recover  the marginal posterior distribution of the {\it Mean Anomaly} physical parameter used in the Keplerian model detailed in Section~S.8  of the  Supplementary Material. The Hamiltonian Monte Carlo sampling is performed using the \texttt{RStan} package \citep{stan_development_team_stan_2019,Rstan_version}. We focus on the {\it Mean Anomaly} because its posterior distribution is expected to be multimodal.  
To obtain a  baseline for comparison, we also applied brute-force numerical integration (at a high computational cost) to obtain the true posterior distribution of the model parameters.  The right panel of Figure~\ref{fig:realdatasampling} compares the marginal posterior distribution of the {\it Mean Anomaly} obtained under the three  different sampling methods. The figure demonstrates that our adaptive Warp-U MCMC sampler more accurately recovers the marginal posterior distribution than PT and HMC. It is also worth noting that PT required substantial tuning, whereas our Warp-U MCMC method did not (the Stan package automatically ran tuning required for HMC).

\noindent \textbf{Estimation of the Bayesian Evidence}.
Next, we compare the performance of stochastic Warp-U bridge estimation with standard bridge sampling estimation and Warp-U bridge estimation.  We applied PT and our adaptive Warp-U MCMC sampler to get the target samples needed by the  estimation methods.  
For fair comparisons, we fix the total number of target density evaluations across the sampling and estimation steps for all methods.

\begin{table}[t]
	\caption{\small{RMSE (and associated SE) when estimating the $\log_{10}$ Bayesian evidence for a planet using bridge sampling, Warp-U bridge estimation, and stochastic Warp-U bridge estimation.}}
	\centering {\small
	\begin{tabular}{r|rr|rr|rr}
		\toprule
		\ &\multicolumn{2}{c}{Bridge} & \multicolumn{2}{|c|}{Warp-U}&\multicolumn{2}{c}{S. Warp-U} \\
		\cmidrule(r){2-7}
		Sampling&RMSE&SE &RMSE&SE&RMSE&SE\\
		\midrule
		PT & $1.144$ & $0.012$& $0.959$& $0.012$& $0.594$&$0.006$  \\
		Warp-U & $0.269$ & $0.004$& $0.277$& $0.005$ &$0.087$ &$0.003$  \\
		\bottomrule
	\end{tabular}}
	\label{tab:Estimation_real}
\end{table}
Following \cite{nelson2018quantifying}, we use the median value of the Bayesian evidence obtained across all the methods investigated  in the Extremely Precise Radial Velocities (EPRV3) Evidence Challenge as the quasi true value (i.e., $\log_{10}(\hat{c}) = -193.71$). Table \ref{tab:Estimation_real} compares the root mean square error (RMSE) for bridge sampling, Warp-U bridge estimation, and stochastic Warp-U bridge estimation, which has the smallest RMSE. Table \ref{tab:Estimation_real} also shows that the RMSE is lower when the samples are from our adaptive Warp-U MCMC sampler than from PT.
The closest estimate to the quasi-true value is  $\log_{10}(\hat{c}) = -193.795$, obtained by stochastic Warp-U bridge estimation using samples from Warp-U MCMC sampler. 
Based on the bias shown in \cite{nelson2018quantifying}, which ignores the variance and therefore represents the methods in \cite{nelson2018quantifying} favorably, this estimator has an RMSE that is comparable to the best methods investigated in the EPRV3 Evidence Challenge.

\section{From Past to Future}\label{sec:literatureReview}

\subsection{A brief Overview of Comparable Methods }\label{sec:litreview}

There is a substantial body of literature on estimating the Bayes factor and the intractable integral in~\eqref{eq:general}. \cite{llorente2023marginal} provide an excellent comprehensive review and  classify existing methods into four  categories: deterministic approximations, density estimation, importance sampling, and vertical representation methods. Bridge sampling is put in the importance sampling category, which includes Chib's method \citep{chib2001marginal} and umbrella sampling \citep{torrie1977nonphysical}.

Many algorithms  have been developed to sample from multi-modal densities, a number of which simultaneously perform sampling and estimation of normalizing constants. A leading example of the latter category of techniques is the Generalized Wang-Landau (GWL) algorithm proposed by \citet{liang2005generalized}, which is an energy based adaptive importance sampling method. The multi-stage approach used in our adaptive method was inspired by the GWL algorithm, and earlier adaptive importance sampling strategies such as \cite{liang2002dynamically}, \cite{berg1991multicanonical}, and \cite{wang2001efficient}. There have been several extensions to the GWL algorithm, including \citet{liang2007stochastic} and \citet{bornn2013adaptive}, but also some concerns about its convergence properties.  \citet{jacob2014wang} showed that only some variations reach the so-called flat histogram convergence criterion in finite time, whereas other variations do not. Furthermore, \citet{wang2022warp} illustrated that the GWL normalizing constant estimator is sometimes inefficient, and the alternative strategy of applying Warp-U bridge estimation to the GWL draws (after weighted resampling)  can substantially reduce RMSE (for fixed computational resources). 

Indeed, although it is conceptually appealing to combine sampling and estimation in a single step, existing techniques for performing these tasks separately are in some ways more developed. 
Some existing algorithms also apply the idea of transporting the mass of the target density, e.g., \cite{parno2018transport}  constructs transport maps to match the target distribution and a reference distribution for more efficient  Metropolis-Hasting proposal. 
\cite{pompe2020framework} proposed a sampling method that begins by finding the modes of the target distribution via optimization;  then, based on this knowledge of the mode locations, they augment the parameter space with a mode index and generate samples via a combination of local moves and mode-jumping moves. Their mode-jumping moves are based on the Metropolis-Hastings algorithm, whereas in our algorithm mode-jumping is achieved via Warp-U transformations.  Furthermore, \cite{tak2018repelling} proposed a repelling attracting Metropolis-Hastings algorithm for exploring multi-modal distributions, by purposefully making move to low density regions before moving back to high density regions.

Perhaps the best known general strategy for sampling from multi-modal densities is parallel tempering \citep{geyer1991markov}. Recent studies \citep{syed2021parallel, syed2022non, surjanovic2022parallel} proposed more efficient version parallel tempering, focusing on its integration with a variational reference distribution, the implementation of flexible annealing paths, and the adoption of non-reversible communication schemes. 
There are also  estimation strategies based on the tempered posterior, which encompass methods such as the stepping stone method \citep{xie2011improving},  annealed importance sampling \citep{neal2001annealed} and  generalized thermodynamic integration \citep{llorente2023target}. In our simulation studies, we find that our proposed sampler is computationally more efficient than parallel tempering because it always accepts inter-mode proposals. While Warp-U sampling does not universally outperform parallel tempering or other methods based on a tempered target distribution, our findings suggest it is a viable alternative with competitive computational cost and implementation effort, showing potential for greater efficiency, especially for high-dimensional multi-modal target distributions with varying mode variances.


\subsection{Limitations and Further Work}\label{sec:further}


A  limitation of our  Warp-U  sampler is its potential inefficiency in the presence of highly isolated modes. This inefficiency arises partly from the time required to identify the modes, and partly from the need for the mixture density to adequately cover all high-density regions of the target distribution to facilitate mode exploration. Sampling from isolated modes with unknown locations remains a universal challenge, particularly in high-dimensional settings. While our sampler has demonstrated remarkable mixing properties in the high-dimensional simulation, its performance relies on the availability of pre-identified modes. On the other hand, in general, it is unclear whether trying to develop samplers that can find modes is a wiser strategy than using optimizers to find modes before applying a sampling method, because optimizers are inherently more suited to finding modes.

Another key requirement for the proposed sampler is that the chosen mixture distribution must accurately approximate the high-dimensional target distribution. In this work, we initialize the mixture distribution using existing variational inference techniques. A promising research direction is to explore mixture distributions with structured covariance matrices, such as low-rank, banded, or sparse precision matrices. These structures can improve computational efficiency in both the initialization of the mixture distribution and the Warp-U sampling process. When there is a non-ignorable discrepancy between the mixture distribution and the target distribution, combining parallel tempering with the Warp-U sampler can be a viable solution. Our simulation studies indicate that the combined sampler can outperform PT alone in such scenarios.  	

Another promising avenue is to incorporate ideas from diffusion samplers \citep{ho2020denoising} to bridge the gap between the mixture and target distributions, leading to transformations with additional injected  random component $\vepsilon$, i.e., non-deterministic Warp-U transformations of the form $\sF_{\zeta}(\vtheta, \vepsilon)$.  Lastly, the neural ODEs approach introduced in Section \ref{sec:neural} offers substantially more flexibility and the numerical results in Section \ref{sec:simulationstudies} demonstrate that it is a promising direction. The key challenge for that method is trading off accurate target approximation with the cost of training the neural network.


There are also some challenges for stochastic Warp-U bridge sampling estimation. For example, when allocating the samples to different components in Steps \ref{SWB_divide_3}-\ref{SWB_divide_4} of Algorithm~\ref{alg:stochasticBridge}, there may be some components that have very few samples, which will lead to high variance of the bridge sampling estimator for those components. To address this problem, future work could develop a restriction to ensure that each component has a minimum number of samples, or by adaptively merging components to ensure sufficient mass for each component.  


Last but not least, much needs to be done to address the biggest theoretical gap in our current article.  That is,  our theoretical comparisons  in Section \ref{sec:theory}  regarding asymptotic variance and precision per CPU second are under the assumption of i.i.d. draws, which is clearly violated by our algorithm and other MCMC methods  used in this article. More painstaking efforts to extend the results to dependent draws are certainly possible, but the more ideal approach is to implement Warp-U sampler perfectly (which {\it would} produce i.i.d. draws), in the sense of the perfect sampling made possible by the seminal work of \cite{propp1996exact}.  Unfortunately, implementing perfect sampling is typically a daunting task, and often impractical; see \cite{craiu2011perfection} for an overview and discussion.   A more pragmatic approach is to develop 
unbiased Warp-U sampler based on the recent work of \cite{jacob2020unbiased}; also see \cite{craiu2022double,wang2023unbiased}.


\bigskip

\spacingset{1.25}

\if1\blind
{
\begin{center}
	{\large\bf Acknowledgements}
\end{center}
We thank the Associate Editor and two reviewers for constructive comments to help us improve the quality of the work significantly.  We thank Yves Atchade and Pierre Jacob for very helpful comments, and US NSF for partial financial support (to XLM). 

} \fi


\begin{center}
{\large\bf SUPPLEMENTARY MATERIAL}
\end{center}

\begin{description}

\item[Supplementary File:]  The supplementary pdf file contains technical proofs and additional numerical results.

\end{description}

\bibliographystyle{apalike}
\bibliography{reference}

\end{document}

%% file: notation.tex
\newcommand{\bbR}{\mathbb{R}}

\newcommand{\sH}{\mathcal{H}}

\newcommand{\sJ}{\mathcal{J}}

\newcommand{\sF}{\mathcal{F}}
\newcommand{\sG}{\mathcal{G}}

\newcommand{\sT}{\mathcal{T}}

\newcommand{\vd}{\boldsymbol{D}}

\newcommand{\vv}{\boldsymbol{v}}

\newcommand{\valpha}{\boldsymbol{\alpha}}
\newcommand{\vtheta}{\boldsymbol{\theta}}

\newcommand{\vmu}{\boldsymbol{\mu}}

\newcommand{\vepsilon}{\boldsymbol{\epsilon}}
\newcommand{\vzero}{\mathbf{0}}
\newcommand{\vxi}{\boldsymbol{\xi}}
\newcommand{\veta}{\boldsymbol{\eta}}

\newcommand{\mI}{\mathbf{I}}

\newcommand{\mS}{\mathbf{S}}

\newcommand{\mSigma}{\boldsymbol{\Sigma}}

\newcommand{\Expect}{\mathbb{E}}

\newcommand{\mud}[1]{\,\mu(\mathrm{d} #1)}

\def\trans{^\mathsf{T}}

\newcommand{\target}{\pi}

\usepackage{color}